\documentclass{article}

\usepackage{lineno}

\usepackage{amsmath}
\usepackage{bbm}
\usepackage{relsize}
\usepackage[plain]{fullpage}
\usepackage{authblk}
\usepackage{graphicx}
\usepackage{amsfonts}
\usepackage{float}

\newcommand{\parderiv}[2]{{ \frac{\partial #1}{\partial #2} }}
\newcommand{\deriv}[2]{{ \frac{\textnormal{d} #1}{\textnormal{d} #2} }}

\begin{document}

\title{A stochastic nonlinear model of the dynamics of actively Q-switched lasers}

\author[1]{Lukas Tarra}
\author[1]{Andreas Deutschmann-Olek}
\author[2]{Vinzenz Stummer}
\author[2]{Tobias Fl\"ory}
\author[2]{Andrius Baltuska}
\author[1,3]{Andreas Kugi}

\affil[1]{\footnotesize Automation and Control Institute, TU Wien, Gu\ss hausstra\ss e 27-29, 1040 Vienna, Austria}
\affil[2]{\footnotesize Photonics Institute, TU Wien, Gu\ss hausstra\ss e 27-29, 1040 Vienna, Austria}
\affil[3]{\footnotesize Center for Vision, Automation \& Control, Austrian Institute of Technology, Giefinggasse 4, 1210 Vienna, Austria}






\maketitle

\begin{abstract}
\noindent In this paper, we present a novel stochastic and spatially lumped multi-mode model to describe the nonlinear dynamics of actively Q-switched lasers and random perturbations due to amplified spontaneous emission. This model will serve as a basis for the design of (nonlinear) control and estimation strategies and thus a high value is set on its computational efficiency. Therefore, a common traveling-wave model is chosen as a starting point and a number of model-order reduction steps are performed. As a result, a set of nonlinear ordinary differential equations for the dynamic behavior of the laser during a switching cycle is obtained. A semi-analytic solution of these differential equations yields expressions for the population inversion after a switching cycle and for the output energy, which are then used to formulate a nonlinear discrete-time model for the pulse-to-pulse dynamics. Simulation studies including models with different levels of complexity and first experimental results demonstrate the feasibility of the proposed approach. 
\end{abstract}

\section{Introduction}
\label{section:introduction}
Nanosecond laser pulses are widely used in areas such as materials processing \cite{materials_processing_application}, spectroscopy \cite{spectroscopy_application}, plasma research \cite{plasma_application} and medical applications \cite{medical_application}, where high pulse energies are required, and also in lidar systems \cite{lidar_application}. The most established technology for achieving such pulses is active Q-switching. Since high pulse repetition rates are desirable for many applications \cite{desire_for_high_rep}, current research efforts are focussed on building actively Q-switched lasers which can operate at up to 1 MHz. At such frequencies and usual pump powers, large variations in output pulse energies which, in most cases, are detrimental to the intended application can ultimately damage optical elements inside the cavity. These energy variations are primarily the result of two distinct effects.
Firstly, spontaneous emission is a stochastic process with known properties \cite{theory_stochastic_effect} which influences the shape of a pulse as well as its total energy (\cite{experiment_stochastic_effect}, Figs. 12 and 13). Secondly, it is already known from regenerative amplifiers \cite{deutschmann} that the pulse-to-pulse dynamics of pumping and inversion depletion exhibits instability regions where pulse energies are elements of nontrivial limit cycles: strong pulses which entail substantial population depletion and therefore strongly weaken the subsequent pulse's gain are followed by weak pulses which allow the populations to recover etc. In the case of actively Q-switched lasers, simulations (\cite{simulation_instability}, Fig.13) and experiments \cite{instability_experiment} have hinted at similar instabilities which ultimately yield (deterministically) chaotic behavior. Hardware-based techniques have been applied to mitigate stochastic \cite{prelasing_state_of_the_art} and dynamic effects \cite{Fluorescence_Feedback} but limit the flexibility of the system. These limitations could be avoided by employing active feedback control strategies as in \cite{deutschmann}, which in turn requires accurate modeling of the system's behavior. 
A review of existing mathematical models of actively Q-switched lasers can be found in \cite{model_review}. 
These models always involve some form of rate equation(s) for the lasing population densities and hence the optical gain of the active medium. A second set of equations then describes the evolution of optical energy within the cavity depending on the current gain and losses. 
On a more detailed level, we will differentiate mathematical models along three mutually independent key characteristics:
\begin{enumerate}
\item Spatially distributed or spatially lumped: In the former approach, population densities and optical powers are assumed spatially dependent (especially along the axis of beam propagation), while the latter disregards power transport phenomena and concentrates populations and powers into one point, see, e.g., \cite{point_model_2004}. Studies have shown \cite{point_vs_travelling, model_review} that spatially distributed modeling achieves superior accuracy especially for long active media.
\item Multi-mode or single-mode: Multi-mode models take into account the spectrally varying and potentially competing gain/loss characteristics of different longitudinal optical cavity modes as pointed out e.g. in \cite{multimode_model}. Several techniques have been invented \cite{mode_selection_review} to select a single longitudinal mode and avoid pulse energy variations resulting from mode-hopping in a mode competition. Single-mode models assume a single effective gain and loss and represent radiation by one longitudinal mode, preventing the simulation of spectral selection with the benefit of shorter overall simulation times.
\item Stochastic or deterministic: Contributions of spontaneous emission to the optical power are essential especially for the modeling of self-seeded lasers as they define the initial intracavity power level. If said contributions are represented by stochastic processes (deterministic quantities), we will call a model stochastic (deterministic). To the authors' knowledge, the stochastic nature of spontaneous emission has so far not been systematically considered in the literature on Q-switched lasers.
\end{enumerate}
The current state of the art is the deterministic and spatially distributed multi-mode approach of traveling-wave modeling as used recently, e.g., in \cite{stateofart_2021, stateofart_2020_2, stateofart_2018}. Traveling-wave models feature spatially dependent rate equations as well as power transport equations for every longitudinal mode. The complex behavior resulting from the coupled nonlinear partial differential equations is useful for simulation studies, but renders an analysis of the pulse-to-pulse dynamics difficult. However, its superior accuracy to other existing models also makes traveling-wave modeling a good starting point for this paper.
Using significant simplifications, Degnan \cite{Degnan} started off with a deterministic and spatially lumped single-mode model and arrived at a transcendental relation between the population inversions before and after a pulse. This is the first and, to the authors' knowledge, only explicit expression for pulse-to-pulse dynamics so far. Unfortunately, the assumption in this paper that the energy build-up always goes to the threshold (the point where the power gains equal the losses and no more amplification occurs) is no longer valid at higher pulse repetition rates, which makes this work not applicable to our problem. 
Therefore, we first present a first principles-based model of actively Q-switched lasers which employs a stochastic generalization of traveling-wave modeling where spontaneous emission is represented by a Bose-Einstein distributed noise term. For the inhomogeneous gain, this paper uses the exemplary case of Nd:YAG active media. Taking the general model as a reference, a number of successive simplification steps are then performed to derive a similarly accurate, but spatially lumped continuous-time stochastic multi-mode model and a discrete-time model which explicitly describes the pulse-to-pulse dynamics.  
Such a discrete-time model is beneficial for three reasons: Firstly, instead of computationally demanding and time-consuming numerical simulations for the complete Q-switching cycle, a single (semi-)analytic expression is given for the population inversion and output pulse energy, respectively. Secondly, the model and its derivation throw light on the question how existing models from the literature are connected, as we will discuss in Section \ref{section:model_comparison}. Finally, the main motivation for this model is the fact that it serves as a theoretical basis for the design of control and estimation algorithms to actively stabilize the pulse-to-pulse dynamics and to suppress stochastic influences on the intracavity power.
Thus, Section \ref{section:general model} first establishes the general framework and the fundamental equations of the model. The step-by-step simplifications are then performed in Section \ref{section:model_simplification}. In Section \ref{section:results}, we show simulation results which illustrate the stochastic and dynamic contributions leading to energy fluctuations. Based on these simulation results, agreement with first measurements and theoretical results on pulse statistics and modeling errors introduced by the various simplification steps are assessed in Section \ref{section:results} as well. The paper concludes with a summary of the presented work and an outlook in Section \ref{section:conclusion}.

\section{General model}
\label{section:general model}
\subsection{Rate and transport equations}

\begin{figure}
\begin{center}
\includegraphics[width=0.7\textwidth]{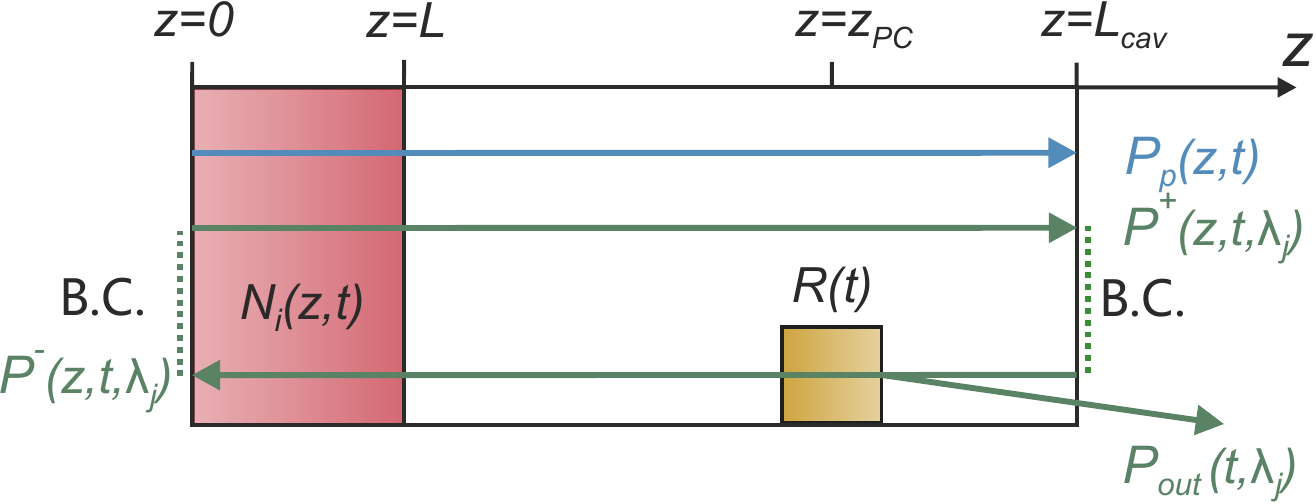}
\caption{A sketch of the cavity, active medium (red area), Q-switch (Pockels cell, yellow area) and forward- and backwards-running pump (blue) and signal (green) powers. The coupling of the two beam propagation directions as well as all other optical components are accounted for via the boundary conditions B.C. (green dots).}
\label{fig:Cavity_Skizze}
\end{center}
\end{figure}

\begin{figure}
\begin{center}
\includegraphics[width=0.35\textwidth]{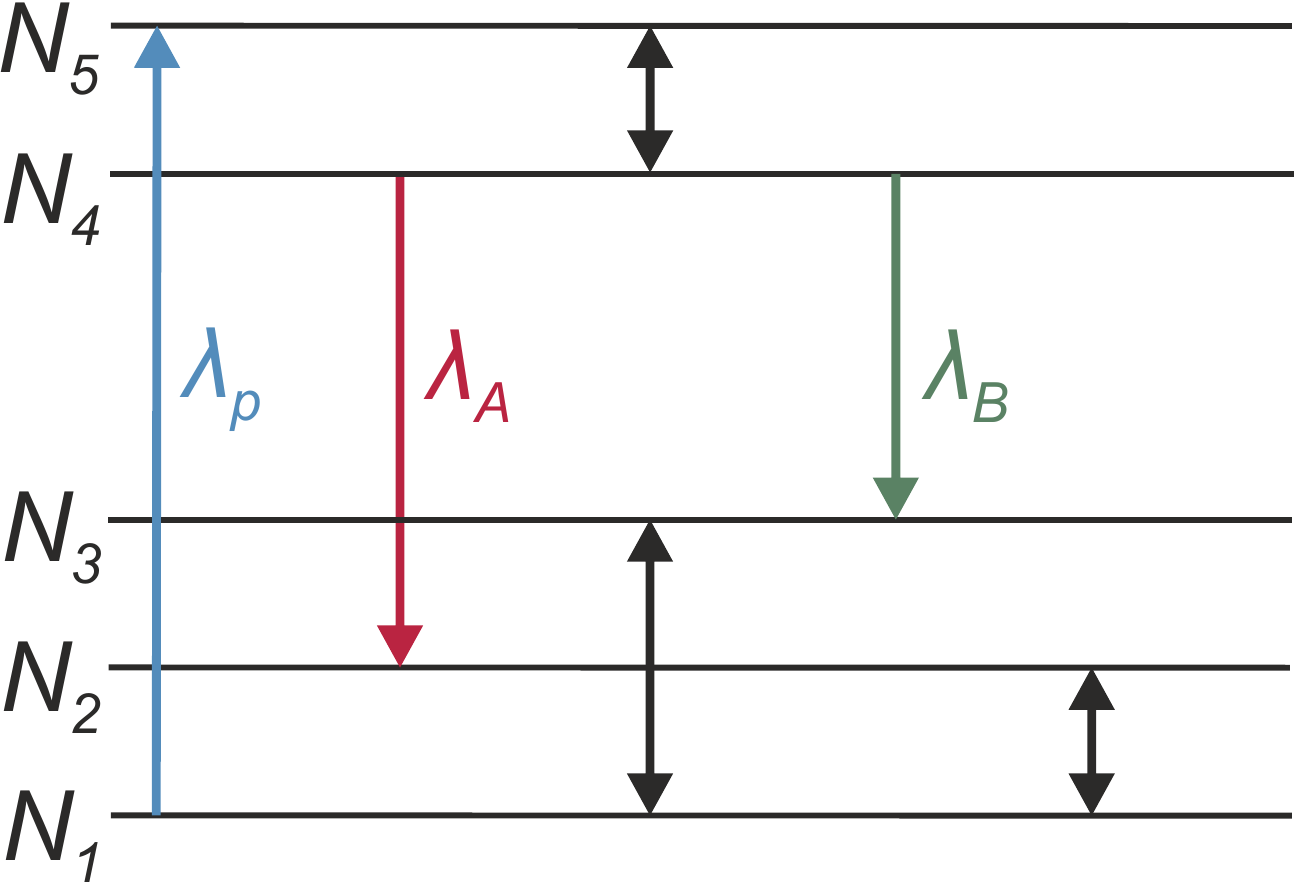}
\caption{The five energy occupation levels that constitute our model and their transitions: pumping (blue) and two lasing transitions (red, green). The black arrows stand for nonradiative thermal transitions.}
\label{fig:energy-scheme}
\end{center}
\end{figure}

Contrary to regenerative amplifiers, the cavities of Q-switched lasers are constantly filled with continuous-wave optical energy, while the pulse is created via the Q-switching process and is usually of greater duration than the cavity round-trip time.  
Throughout this work, we will make the usual assumption that no two laser modes of different wavelengths are mutually coherent and thus that no interference phenomena such as mode-locking occur. Therefore, radiation will be represented by powers instead of electric fields.
Our starting point will be the (multi-mode) traveling-wave equations, where we use the example of a forward pumped Nd:YAG crystal within a linear (Fabry-Perot) cavity. A sketch of the system is drawn in Fig. \ref{fig:Cavity_Skizze}. The spatial coordinate inside the cavity along which light propagates is defined as $z$-direction, and the active medium with population densities $N_i(z,t)$ resides within the interval $[0,L]$. Population densities and optical powers will always be assumed to be transversally uniform, which is why there will be no spatial dependences other than on $z$. At $z = z_{PC}$, a Pockels cell with a reflection coefficient $R(t)$ performs the Q-switching. All optical components apart from the active medium and the Q-switch determine the boundary conditions at either end of the cavity, which are specified in Section \ref{section:boundary_conditions}. 
For Nd:YAG active media, the most interesting lasing transitions occur at wavelengths $\lambda_A$ = 946 nm and $\lambda_B$ = 1064 nm. Their technological relevance is due to a small quantum defect and hence high theoretical efficiency in the former and a large emission cross-section in the latter \cite{Nd:YAG}, which renders the 1064 nm transition the dominant one. Pumping can be achieved at $\lambda_p$ = 809 nm from the ground state to an elevated state, which quickly decays to the upper state for both lasing transitions considered. Thus, our full model of the active medium effectively consists of five population densities as visualised in Fig. \ref{fig:energy-scheme}.
The rate equations for the population densities $N_i(z,t)$ with the mentioned underlying energy level scheme and a total of $M$ longitudinal cavity modes read \cite{model_review}
\begin{subequations} \label{equ:rate_equations}
\begin{align}
	\label{popeqn1}&\parderiv{N_5(z,t)}{t} = -\gamma_{54} N_5(z,t) + \gamma_{45} N_4(z,t) + \frac{ \lambda_{p} \sigma_{p}}{h c A_p} P_p(z,t) \left(N_1(z,t) - N_5(z,t)\right)\\
\label{popeqn2} & \parderiv{N_4(z,t)}{t} = \gamma_{54} N_5(z,t) - \gamma_{45} N_4(z,t) - (\gamma_{42} + \gamma_{43}) N_4(z,t) - \sum_{j=1}^M \frac{ \lambda_j}{h c A_s} \times\\
& \times  \left(P^{+}(z,t,\lambda_j) + P^{-}(z,t,\lambda_j)\right)\Big(\sigma_A(\lambda_j) \left(N_4(z,t) - N_2(z,t)\right) +\sigma_B(\lambda_j) \left(N_4(z,t) - N_3(z,t)\right)  \Big) \notag \\
	\label{popeqn3}\begin{split}& \parderiv{N_3(z,t)}{t} = -\gamma_{31} N_3(z,t) + \gamma_{13} N_1(z,t) + \gamma_{43} N_4(z,t) \\
	& +  \sum_{j=1}^M \frac{ \lambda_j}{h c A_s} \left(P^{+}(z,t,\lambda_j) + P^{-}(z,t,\lambda_j)\right)\Big(\sigma_B(\lambda_j) \left(N_4(z,t) - N_3(z,t)\right)  \Big) \end{split} \\ 
	\label{popeqn4}\begin{split} &\parderiv{N_2(z,t)}{t} = -\gamma_{21} N_2(z,t) + \gamma_{12} N_1(z,t) + \gamma_{42} N_4(z,t)  \\
	& +  \sum_{j=1}^M \frac{ \lambda_j}{h c A_s} \left(P^{+}(z,t,\lambda_j) + P^{-}(z,t,\lambda_j)\right)\Big(\sigma_A(\lambda_j) \left(N_4(z,t) - N_2(z,t)\right)  \Big) \ ,  \end{split}
	\end{align}
	\end{subequations}
with relaxation rates $\gamma_{ij}$ for the populations and absorption and induced emission terms due to the pump power $P_p(z,t)$ as well as forward- and backward-running lasing powers $P^\pm(z,t,\lambda_j)$. Wavelengths $\lambda_j$ belong to a discrete spectrum of cavity modes. The cross-sectional areas of the pump and the signal beams are given by $A_p$ and $A_s$, respectively. The absorption/emission cross-sections (which we set equal for simplicity) are denoted by $\sigma_A(\lambda_j)$ and $\sigma_B(\lambda_j)$ for the transition wavelengths $\lambda_A$ and $\lambda_B$, respectively. They determine the spectral lineshapes of the two transitions (which in general may overlap). $N_{dop}$ is the crystal's total Neodymium dopand density, and $N_1(z,t)$ can be calculated from the other population densities via the algebraic equation $N_1(z,t) = N_{dop} - N_5(z,t) - N_4(z,t) - N_3(z,t) - N_2(z,t)$. For setups such as cladding-pumped lasers, one also needs to include appropriate overlapping factors $\Gamma$, see, e.g., \cite{model_review}. \\
Because active media usually have very similar refractive indices at comparable wavelengths $\lambda_A$, $\lambda_B$, and $\lambda_p$, the group velocities of radiation in this spectral range are all approximately equal to a common value $v$. The power transport equations can then be written as (\cite{laser_electronics}, Cha. 8)
\begin{subequations} \label{equ:power_transport_equations}
\begin{align}	
& \frac{1}{v}\parderiv{P_p(z,t)}{t} + \parderiv{P_p(z,t)}{z}  = \sigma_{p} P_p(z,t) \left( N_5(z,t)-N_1(z,t) \right)-\alpha_p P_p(z,t)\label{powereqn1}\\
 & \frac{1}{v}\parderiv{P^\pm(z,t,\lambda_j)}{t} \pm \parderiv{P^\pm(z,t,\lambda_j)}{z} = P^\pm(z,t,\lambda_j)\Big( \sigma_A(\lambda_j) \left( N_4(z,t)-N_2(z,t) \right) \ +\label{powereqn2} \\
&  \sigma_B(\lambda_j) \left( N_4(z,t)-N_3(z,t) \right) \Big)  -\alpha(\lambda_j) P^\pm(z,t,\lambda_j)  +\alpha_{RS} P^\mp(z,t,\lambda_j) + \frac{2hc^2}{\lambda_j^3} \Delta \lambda_j \frac{\Delta\Omega}{4\pi} \zeta(z,t,\lambda_j) \ ,  \notag
\end{align}
\end{subequations}
where the first terms on the right-hand sides stand for absorption and induced emission due to interaction with the population densities $N_i(z,t)$. The second terms represent effective losses, for example diffraction, with the coefficients $\alpha_p$ and $\alpha(\lambda_j)$, respectively, and $\alpha_{RS}$ is the Rayleigh backscattering coefficient (capture factor included). The final term on the right-hand side of (\ref{powereqn2}) accounts for spontaneous emission, where $2hc^2/\lambda_j^3 \Delta \lambda_j = 2h\nu_j \Delta \nu$ is the contributed power of spontaneously emitted photons of 2 independent polarisations belonging to a mode $\lambda_j$ with a spectral spacing $\Delta \lambda_j$ between modes into a solid angle $\Delta\Omega$ where photons are kept inside the cavity \cite{ASE_spectral}. Moreover, $\zeta(z,t,\lambda_j)$ is random in nature and describes the number of photons of wavelength $\lambda_j$ spontaneously emitted in the region $[z,z+\textnormal{d}z)$ at time $t$ per length increment $\textnormal{d}z$. The term will be further discussed in Section \ref{section:stochastic_seed}. 

\subsection{Cavity boundary conditions}
\label{section:boundary_conditions}
As can be seen in Fig. \ref{fig:Cavity_Skizze}, optical elements (whose effective round-trip cavity efficiency is $\eta(\lambda_j)$) bring along that the pump and beam powers are subject to boundary conditions (B.C.)
\begin{subequations} \label{equ:boundary_conditions}
 \begin{align}
&P_p(0,t) = P_p(t) \label{equ:B_C_1}  \\
&P^+(0,t,\lambda_j) = \sqrt{\eta(\lambda_j)} P^-(0,t,\lambda_j) \label{equ:B_C_2} \\
&P^-(L_{cav},t,\lambda_j) = \sqrt{\eta(\lambda_j)} P^+(L_{cav},t,\lambda_j) \ . \label{equ:B_C_3}
\end{align}
\end{subequations}
For $R(t)\neq 1$, the Pockels cell introduces a discontinuity at $z_{PC}$ given by
\begin{subequations}
\begin{align}\label{equ:Q_switch}
&P^-(z_{PC}-\textnormal{d}z ,t+ \textnormal{d}z / v,\lambda_j) = R(t) P^-(z_{PC},t,\lambda_j) \ ,
\end{align}
with the resulting output power
\begin{align}\label{equ:P_out_full}
& P_{out}(t,\lambda_j) =  \left(1-R(t)\right) P^-(z_{PC},t,\lambda_j) \ .
\end{align}
\end{subequations}
From this and the Q-switching frequency $f_{switch}$, we get the $m$-th output pulse energy as
\begin{align} \label{equ:E_out}
E_{m} = \sum_{j=1}^M \int_{(m-1)/f_{switch}}^{m/f_{switch}} P_{out}(t,\lambda_j) \ \textnormal{d}t \ .
\end{align}

\subsection{Stochastic seeding via spontaneous emission}
\label{section:stochastic_seed}
First, we define the total scattering cross-section
\begin{align}\label{equ:total_cross_section}
\sigma(\lambda_j):=\sigma_A(\lambda_j) + \sigma_B(\lambda_j) \ .
\end{align} 
We will use it for the ensemble mean of $\zeta(z,t,\lambda_j)$,
\begin{align}\label{equ:inverse_mean_free_path}
\mathbb{E}(\zeta(z,t,\lambda_j)) = \sigma(\lambda_j) N_4(z,t) \ ,
\end{align}
which corresponds to the inverse mean free path. It can be shown quantum-mechanically for a highly simplified model \cite{Bose_Einstein_derivation}, where for instance, all longitudinal modes experience identical gains, that amplified spontaneous emission in $K$ independent modes belonging to one bosonic excitation state follows the $K$-fold degenerate Bose-Einstein distribution, which for large $K$ approaches a Poisson distribution and for low $K$ is significantly broader \cite{Bose_Einstein_Poisson}. Since we simulate every cavity mode individually, we can set $K=1$ and therefore propose that the probability of $n$ photons of wavelength $\lambda_j$ being spontaneously emitted within $[0,z]$ at time $t$ follows a single-mode Bose-Einstein distribution
\begin{align}\label{equ:Bose_distribution}
&\mathbb{P}[\mathfrak{C}(z,t,\lambda_j) = n] = \left( \frac{\mu(z,t,\lambda_j)}{1 + \mu(z,t,\lambda_j)} \right)^n \frac{1}{1 + \mu(z,t,\lambda_j)} \ ,
\end{align}
with mean
\begin{align}\label{equ:mu_intensity_definition}
\mu(z,t,\lambda_j) := \sigma(\lambda_j)\int_0^z N_4(z',t) \ \textnormal{d}z' \ .
\end{align}
Hence, the number of photon counts,
\begin{align} 
\mathfrak{C}(z,t,\lambda_j) := \int_0^z \zeta(z',t,\lambda_j) \ \textnormal{d}z' \ ,
\end{align}
is an inhomogeneous jump process following the probability distribution (\ref{equ:Bose_distribution}).  \\
\noindent The dominant mode $\lambda_j$ of a pulse arises from a competition among modes around the maximum of $\sigma_{A}(\lambda_j)$ or $\sigma_{B}(\lambda_j)$ due to the greatest amount of spontaneously emitted photons in the early phase of the energy build-up.  Bragg gratings and other wavelength-selective elements inside the cavity can shape $\eta(\lambda_j)$ in the boundary conditions (\ref{equ:B_C_1}-\ref{equ:B_C_3}) such that the setup exhibits a few or many significant modes. As a consequence, the experimental output pulses will be approximately Bose-Einstein or Poisson distributed, as was shown in \cite{Bose_Einstein_Poisson}.

\section{Model simplification}
\label{section:model_simplification}

\subsection{Transport equation solution \& total rate equations}
\begin{figure}
\begin{center}
\includegraphics[width=0.7\textwidth]{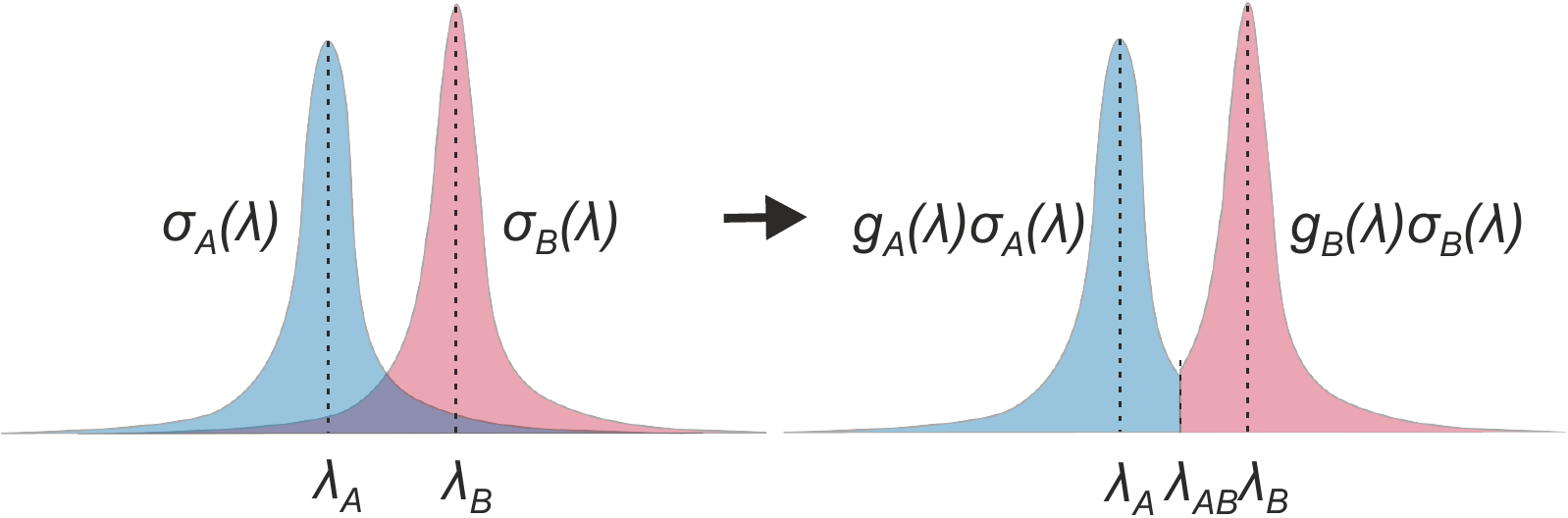}
\caption{The overlap between lineshapes of scattering cross-sections $\sigma_A(\lambda)$ and $\sigma_B(\lambda)$ is eliminated by lineshape separators $g_A(\lambda)$ and $g_B(\lambda)$.}
\label{fig:Lineshapes}
\end{center}
\end{figure}

For control-related tasks, numerically solving the coupled rate (\ref{popeqn1}-\ref{popeqn4}) and power transport (\ref{powereqn1},\ref{powereqn2}) equations with the boundary conditions (\ref{equ:B_C_1}-\ref{equ:Q_switch}) is not feasible as systematically investigating the pulse-to-pulse dynamics based solely on simulations is difficult. To eliminate spatial dependences, we introduce the total populations
\begin{align}\label{equ:N_tot}
N_i^{tot}(t):=\int_0^L N_i(z,t) \ \textnormal{d}z
\end{align}
and make a number of simplifications which allow to express the dynamical behavior of the laser in terms of these new variables.\\
We start the model simplification by choosing frames $t \rightarrow t \mp z/v$ in (\ref{powereqn1},\ref{powereqn2}) which move with a point of the pump or signal beam. Additionally, we make the following two assumptions:\\
 (A1.) When the forward-running (the backward-running) laser beam travels through the active medium starting at $t_k$, the population densities and the lasing powers are assumed to be temporally constant, i.e. $ N_i(z,t) = N_i(z,t_k)$ and $P^{\mp}(z,t) = P^{\mp}(z,t_k)$ for $t \in [t_k,t_k+L/v)$ in (\ref{equ:power_transport_equations}).  Moreover, for the Rayleigh backscattering contribution $\alpha_{RS}P^{\mp}(z,t,\lambda_j)$, the dependence of $P^{\mp}$ on $z$ will be neglected, i.e. $P^-(z,t,\lambda_j)=P^-(L,t,\lambda_j)$ and $P^+(z,t,\lambda_j)=P^+(0,t,\lambda_j)$.
The transport phenomena can therefore be described completely in terms of the forward- and backward-moving frames $t \rightarrow t \mp z/v $.\\
(A2.) Photons that are spontaneously emitted or backscattered at $z \in [0,L]$ join the forward-running (the backward-running) beam at $z=L$ ($z=0$). In the supplemental document, we provide the formulae representing this assumption. \\
With initial conditions given by the incident powers $P_p(0,t_k) = P_p(t_k)$ and $P^\pm(\frac{L\mp L}{2},t_k,\lambda_j)$ for the pump and signal beams before transmission, respectively, and separating each two channels' frequencies by one inverse cavity round-trip time, i.e. $\Delta \nu = 1/t_{RT}$, we show in the supplemental document that the power transport equations (\ref{equ:power_transport_equations}) can be solved analytically provided that the assumptions (A1.) and (A2.) hold. The solutions, i.e. the outgoing powers corresponding to any incoming ones, read
\begin{subequations}
\begin{align}
&P_p(L,t_k+L/v) =  \exp \Big( \sigma_{p} (N_5^{tot}(t_k) - N_1^{tot}(t_k)) -\alpha_pL \Big) P_p(0,t_k) \label{P_inout1}\\ 
\begin{split}
 &P^\pm\left(\frac{L\pm L}{2},t_k+L/v,\lambda_j\right) =   2 \frac{hc}{\lambda_j t_{RT}} \frac{\Delta \Omega}{4\pi} \mathfrak{C}(t_k,\lambda_j) + \alpha_{RS}L P^\mp\left(\frac{L\pm L}{2},t_k,\lambda_j\right) \ + \label{P_inout2} \\
&  \ \ \ \ \ \ \ \ \ \ \ \ P^\pm\left(\frac{L\mp L}{2},t_k,\lambda_j\right) \exp \Big( \sigma_A(\lambda_j) \left(N_4^{tot}(t_k) - N_2^{tot}(t_k)\right) + \\
&  \ \ \ \ \ \ \ \ \ \ \ \ \sigma_B(\lambda_j) \left(N_4^{tot}(t_k) - N_3^{tot}(t_k)\right) -\alpha(\lambda_j)L \Big) \ , 
\end{split}
\end{align}
\end{subequations}
where the spontaneously emitted photon count $ \mathfrak{C}(t_k,\lambda_j) := \mathfrak{C}(L,t_k,\lambda_j)$ is a random number with distribution (\ref{equ:Bose_distribution}) and mean $\mu(L,t_k,\lambda_j)$ according to (\ref{equ:mu_intensity_definition}), i.e. the value the counting process has reached between $z=0$ and $z=L$.
As a next step, we want to find rate equations for the total populations defined by (\ref{equ:N_tot}). To this end, we make use of the following assumptions:\\
(A3.) When calculating the depletion of population inversion due to lasing powers or its increase due to the pump power, dependence on $z$ of the loss terms will be neglected, i.e. $\alpha(\lambda_j)P^{+}(z,t,\lambda_j) = \alpha(\lambda_j)P^{+}(0,t,\lambda_j)$ and $\alpha(\lambda_j)P^{-}(z,t,\lambda_j) = \alpha(\lambda_j)P^{-}(L,t,\lambda_j)$ as well as $\alpha_p P_p(z,t) = \alpha_p P_p(0,t)$.\\
(A4.) For the induced emission contributions to the lower populations $N_3$ and $N_2$ in (\ref{popeqn1}-\ref{popeqn4}), spectral overlap of the lineshapes $\sigma_A(\lambda)$ and $\sigma_B(\lambda)$ can be disregarded. \\
 Assumption (A4.) may be implemented using lineshape separator functions $g_A(\lambda)$ and $g_B(\lambda)$ which are either 1 or 0 and switch support at some separation wavelength $\lambda_{AB}$. The effect of lineshape separators can be seen in Fig. \ref{fig:Lineshapes}.
With assumptions (A1.)-(A4.), we show in the supplemental document that the total rate equations read as
 \begin{subequations}
\begin{align}
\begin{split} &\deriv{N_5^{tot}(t)}{t} = -\gamma_{54} N_5^{tot}(t) + \gamma_{45} N_4^{tot}(t) \ + \\
& \ \ \ \ \ \ \ \ \ \ \ \ \ \ \ \ \ \ \  \frac{\lambda_{p}}{h c A_p} \bigg(1 - \exp \Big( \sigma_{p} (N_5^{tot}(t) - N_1^{tot}(t)) -\alpha_pL \Big) -\alpha_pL \bigg) P_p(t) \label{full_fia1} \end{split} \\
 \begin{split}
&\deriv{N_4^{tot}(t)}{t} = - (\gamma_{45} + \gamma_{43} + \gamma_{42}) N_4^{tot}(t) + \gamma_{54} N_5^{tot}(t) - \sum_{j=1}^M G(t,\lambda_j) \end{split} \\
 \begin{split}
 &\deriv{N_3^{tot}(t)}{t} = - \gamma_{31} N_3^{tot}(t) + \gamma_{13} N_1^{tot}(t) + \gamma_{43} N_4^{tot}(t) + \sum_{j=1}^M g_B(\lambda_j) G(t,\lambda_j) \end{split} \\
 \begin{split}
&\deriv{N_2^{tot}(t)}{t} = - \gamma_{21} N_2^{tot}(t) + \gamma_{12} N_1^{tot}(t) + \gamma_{42} N_4^{tot}(t)  + \sum_{j=1}^M g_A(\lambda_j) G(t,\lambda_j) \label{full_fia4}
\end{split} \\
&G(t,\lambda_j):=  \frac{\lambda_j}{h c A_s} \left(P^+(0,t,\lambda_j)+P^-(L,t,\lambda_j) \right) \times \label{equ:G_def}  \\
 & \times \bigg( \exp \Big( \sigma_A(\lambda_j) \left(N_4^{tot}(t) - N_2^{tot}(t)\right) + \sigma_B(\lambda_j) \left(N_4^{tot}(t) - N_3^{tot}(t)\right) -\alpha(\lambda_j)L \Big) -1 + \alpha(\lambda_j)L \bigg) \ , \notag
\end{align}
\end{subequations}
with the algebraic equation $N_1^{tot}(t) = L N_{dop} - N_5^{tot}(t) - N_4^{tot}(t)- N_3^{tot}(t)- N_2^{tot}(t)$ . 

\subsection{State reduction via slow-fast dynamics}

The model so far considers 5 populations of which only 4 are independent due to the algebraic condition involving $N_{dop}$. If the relaxation rate $\gamma_{ik}$ of an excitation state population $N_i$ is very large, the state is depopulated nearly instantly to $N_k$. In that case, which is usually a good approximation for low energy gaps $\Delta E_{ik}$ in the order of phonon energies, we can apply singular perturbation theory to reduce the order of the model.
Assuming that the active medium is at thermal equilibrium (thus introducing a thermalization factor $B_{45} := \exp\left(\frac{-\Delta E_{45}}{k_BT}\right)$) \cite{Boltzmann_distribution} and employing a linear state transform, we show in the supplemental document that the total rate equations (\ref{full_fia1}-\ref{full_fia4}) consist of a slow and a fast dynamic subsystem. The limit of infinitely fast relaxation yields a quasi-stationary model for the parameters
\begin{align}
\gamma_{31} = a_{31} \gamma_{54}, \ \gamma_{21} = a_{21} \gamma_{54}, \  a_{31} > 0, \  a_{21} > 0, \  \gamma_{54} \rightarrow \infty \ , \label{complete_red_limit}
\end{align}
which results in a slow dynamics that can be expressed solely in terms of the total population
\begin{align}
 N(t) :=  N_4^{tot}(t) \ . \label{equ:total_inversion}
\end{align}
In light of (\ref{equ:G_def}), a reasonable choice of the spatially lumped intracavity power can be given by the power directly applying to the active medium, which is given by
\begin{align}
P_j(t) := P^+(0,t,\lambda_j) + P^-(L,t,\lambda_j) \label{equ:total_P_definition} \ .
\end{align}
In the limit (\ref{complete_red_limit}) with (\ref{equ:total_inversion}), (\ref{equ:total_P_definition}) and (\ref{equ:total_cross_section}) and defining $b:=1/(1+B_{45})$ as well as $\gamma:=\gamma_{42} + \gamma_{43}$, we show in the supplemental document that one can obtain from (\ref{full_fia1}-\ref{full_fia4}) the reduced total rate equation
\begin{align}
\begin{split}
\deriv{ N(t)}{t} & =  - b \gamma N(t) - b \sum_{j=1}^M\frac{\lambda_j}{h c A_s}  \bigg( \exp \Big(\sigma(\lambda_j) N(t) -\alpha(\lambda_j)L \Big) -1 + \alpha(\lambda_j)L \bigg) P_j(t)  \\
& + b \frac{\lambda_{p}}{h c A_p} \bigg( 1- \exp \Big(  \sigma_{p} \left((2 B_{45} + 1) N(t) -L N_{dop} \right) -\alpha_pL \Big) -\alpha_pL \bigg) P_p(t)\ .\label{comp_red_rate} \end{split} 
\end{align}

\subsection{Continuous-time nonlinear model (CTNM)}
\label{section:CTNM}

Since we are mainly concerned with the evolution of the intracavity power rather than the transport effects inside the cavity, we will average the transmission through the active medium, the cavity losses and the Q-switch influence over half a cavity round-trip for a linear cavity, as was done in \cite{Degnan}.
For this, the power transport solution (\ref{P_inout2}) and the boundary conditions (\ref{equ:B_C_2},\ref{equ:B_C_3},\ref{equ:Q_switch}) are utilized and for simplicity, $\sqrt{R(t)}$ is applied to both forward- and backward-running beams instead of $R(t)$ to the backward beam. We also assume that amplification and losses also apply to spontaneously emitted or backscattered photons. Then, half a cavity round-trip, i.e. a time increment of $\Delta t := t_{RT}/2 = L_{cav}/v$, affects the powers according to
\begin{align} \label{equ:P_roundtrip}\begin{split}
&P^\pm\left(\frac{L\mp L}{2},t_{k+1},\lambda_j\right) =\sqrt{ \eta(\lambda_j) R(t_k)} \exp\left(\sigma(\lambda_j) N(t_k)-\alpha(\lambda_j)L\right) \times \\
& \times \left( P^\mp\left(\frac{L\pm L}{2},t_k,\lambda_j\right) +   2 \frac{hc}{\lambda_j t_{RT}} \frac{\Delta\Omega}{4\pi} \mathfrak{C}(t_k,\lambda_j) +\alpha_{RS} LP^\mp\left(\frac{L\pm L}{2},t_k,\lambda_j\right) \right) \ . \end{split}
\end{align}
The time derivative of (\ref{equ:total_P_definition}) can be approximated by
\begin{align} \label{approx_P_derivative}
\deriv{P_j(t)}{t} \approx \frac{P_j(t)}{\Delta t} \left( \ln P_j(t_{k+1}) - \ln P_j(t_{k}) \right) \ ,
\end{align}
which with (\ref{equ:total_P_definition}) and (\ref{equ:P_roundtrip}) and assuming that spontaneous emission and Rayleigh backscattering are small effects (i.e. $\ln(1+x) \approx x$) yields an ordinary differential equation which is derived in the supplemental document and reads as
\begin{align}\label{continuous_P_equation}
\deriv{P_j(t)}{t} = \frac{2\sigma(\lambda_j)}{t_{RT}} N(t) P_j(t) - \frac{P_j(t)}{\tau(t,\lambda_j)} +  \frac{ 2}{t_{RT}} \Big( \frac{ 4 hc}{\lambda_j t_{RT}} \frac{\Delta\Omega}{4\pi} \mathfrak{C}(t,\lambda_j) +\alpha_{RS}L P_j(t) \Big) \ ,
\end{align}
where we have defined the loss rate
\begin{align*}
1/\tau(t,\lambda_j) := \frac{1}{t_{RT}} \left( 2\alpha(\lambda_j) L -  \ln\left( \eta(\lambda_j) R(t)\right) \right) \ .
\end{align*}
Again, $ \mathfrak{C}(t,\lambda_j)$ is a Bose-Einstein noise term with expectation value $\mathbb{E}( \mathfrak{C}(t,\lambda_j)) = \sigma(\lambda_j) N(t)$ that can be understood in a Langevin sense. More formally, one could write (\ref{continuous_P_equation}) as a stochastic differential equation involving a renewal-reward process. We show in the supplemental document how the special case of a compound Poisson process with exponentially distributed waiting times and Poisson distributed jumps can be used to model the emission of photons at random times while producing equivalent means and variances after a time increment of $t_{RT}/2$ as the Bose-Einstein distribution from (\ref{equ:Bose_distribution}).
Neglecting time delays and Rayleigh backscattering during half a round-trip, we show in the supplemental document that the output power, originally given by (\ref{equ:P_out_full}), can be obtained from the spatially lumped intracavity power by 
\begin{align}\label{equ:P_out_CTNM}
P_{out}(t,\lambda_j) = \frac{\left( 1-R(t) \right)}{R(t)} \left(  \frac{P_j(t)\exp\left( \sigma(\lambda_j)N(t)-\alpha(\lambda_j)L \right) +  2 \frac{hc}{\lambda_j t_{RT}} \frac{\Delta\Omega}{4\pi} \mathfrak{C}(t,\lambda_j)}{\frac{1}{\sqrt{\eta(\lambda_j)}R(t)} + \exp\left( \sigma(\lambda_j)N(t)-\alpha(\lambda_j)L \right) }  \right),
\end{align}
while the output energy is still given by (\ref{equ:E_out}).\\
Equ. (\ref{continuous_P_equation}) and the reduced total rate equation (\ref{comp_red_rate}) provide a very compact description of the system dynamics during a switching cycle and constitute what we will call the continuous-time nonlinear model (CTNM) of actively Q-switched lasers. 
In order to obtain the population and power after a switching cycle (and hence the pulse-to-pulse behavior of the system) from the CTNM, one would have to solve the coupled, nonlinear differential equations (\ref{comp_red_rate},\ref{continuous_P_equation}), which is not possible without further approximations, as we will discuss in Section \ref{section:DTNM}.

\subsection{Comparison to existing models}
\label{section:model_comparison}
It is an interesting observation that our modeling effort so far bridges the gap between spatially distributed traveling-wave modeling (where we started our investigation) and the more traditional point models which are spatially lumped. From the spatially lumped total rate equation (\ref{comp_red_rate}), the rate equation used in traditional point models can be derived by isolating the pump and relaxation terms, additionally assuming losslessness and non-pumping-saturation, i.e. $\alpha_p =0$ and $LN_{dop} \gg N(t)$, whose solution yields equation (5) in \cite{degnan_experiment}, and assuming low single-transmission gains, i.e. $\exp(\sigma(\lambda_j) N(t) -\alpha L) \approx 1+(\sigma(\lambda_j) N(t) -\alpha L)$. 
This indicates that the extra terms in (\ref{comp_red_rate}) are effects of the spatially distributed nature of the amplification process of the intracavity power as well as of the absorption of the pump power neglected by traditional point models. These modifications have a significant effect on the model's behavior during a switching cycle and affect the pulse-to-pulse dynamics, especially for low to medium switching frequencies.
Furthermore, the power equation used along with a point model rate equation in Degnan's model \cite{Degnan}, which has been used extensively for experimental laser optimisation \cite{degnan_experiment}, can be retrieved from (\ref{continuous_P_equation}) by neglecting the last terms accounting for spontaneous emission and backscattering and disregarding multi-mode behavior. To the best of the authors' knowledge, the model proposed in this paper is the first one on Q-switched lasers where the stochastic nature of spontaneous emission is incorporated.

\subsection{Pulse-to-pulse dynamics}
\label{section:DTNM}

As pointed out in Section \ref{section:introduction}, the output pulse energies of an actively Q-switched laser are subject to fluctuations due to stochastic and dynamic effects. The former mainly come from amplified spontaneous emission, while the latter are a result of unstable pulse-to-pulse dynamics due to a coupling of subsequent pulses. 
Since Q-switching (e.g. a Pockels cell represented by $R(t)$) operates in a cyclic manner, the pulse-to-pulse dynamics can be described by a discrete-time dynamic system where each switching cycle is a discrete event.
If $t_m$ is the time immediately after the $(m-1)$-th pulse has been coupled out, one switching cycle evolves according to $t_{m+1} = t_m + 1/f_{switch}$. We use the $m$-th pulse energy $E_{m}$ given by (\ref{equ:E_out}) and (\ref{equ:P_out_CTNM}) and the abbreviation $N(t_m):=N_m$. While most generally, $N_m$ and $P_{j}(t_m)$ would both be dynamic variables, we can argue that $P_{j}(t_m) \approx 0$ for all $m$ and $j$ since all intracavity power has just been coupled out at those times, which reduces the dimension of the dynamical system from $1+M$ to $1$. With future control applications in mind, typical feedback quantities include the measured pulse energy, i.e. $E_m$ is the considered output quantity. As input quantity, we can choose the intracavity power $P_j(t_m+t_{pump}) =: u_{j,m}$ after the pumping phase and at the beginning of the $m$-th energy build-up. This input quantity is either given by spontaneous emission (uncontrolled case) or by a defined level from external seeding or controlled prelasing \cite{prelasing_state_of_the_art} (controlled case). Thus, defining the pulse-to-pulse and output maps by $f$ and $h$, respectively, the discrete dynamics are given by
\begin{subequations} \label{equ:DTNM_general}
\begin{align}
N_{m+1} &= f(N_m,u_{j,m}) \label{equ:f_cyc_general} \\
E_m &= h(N_m,u_{j,m}) \ . \label{equ:h_cyc_general}
\end{align}
\end{subequations}
These mappings are either found from numerical solutions of the general model discussed in Section \ref{section:general model}, numerical solutions of (\ref{comp_red_rate},\ref{continuous_P_equation}) with (\ref{equ:P_out_CTNM},\ref{equ:E_out}) or an approximate analytic solution. 
Thus, in order to arrive at explicit expressions for (\ref{equ:DTNM_general}), we will separate the switching cycle into its essential sub-processes and approximately solve equations (\ref{comp_red_rate}) and (\ref{continuous_P_equation}) for each sub-process, incorporating additional fitting parameters. The three essential sub-processes of a switching cycle are:
\begin{enumerate}
\item pumping and population relaxation for a duration of $t_{pump}$,
\item energy build-up during the duty cycle $t_{bu} = \textnormal{DC}/f_{switch}$ (DC is the relative duty cycle), and
\item coupling-out of the pulse with pulsewidth $t_{coup} \approx t_{close}$, where $t_{close}$ is the closing time of the cavity.
\end{enumerate} 
The composition of these solutions then constitutes our discrete-time dynamic model of the pulse-to-pulse dynamics.
This semi-analytic procedure can be done in different ways, of which the authors explored a few. We present a detailed derivation in the supplemental document whose results are relatively simple and can be fitted effectively to adjust the dynamics to the observed behavior. 
What renders the procedure especially difficult is a combination of exponential growth underlying the energy build-up and nonlinearities in both (\ref{comp_red_rate}) and (\ref{continuous_P_equation}). The nonlinearities stemming from spatially distributed amplification and absorption are essential to accurately describe the dynamics, especially for low to medium switching frequencies. Therefore, linearisation or omission of loss terms immediately leads to substantial modeling errors.
In the supplemental document, we show that one can find explicit expressions for the pulse-to-pulse dynamics of the form 
\begin{subequations} \label{equ:DTNM}
\begin{align}
N_{m+1} &= f(N_m,u_{j,m}) = f_{3,N} \left( \mathbf{f}_2 \left( \left[ \begin{array}{c} f_1(N_m)\\\sum_{j=1}^M u_{j,m} / M \end{array} \right] \right)  \right) \\
E_{m} &= h(N_m,u_{j,m}) = f_{3,E} \left( \mathbf{f}_2 \left( \left[ \begin{array}{c} f_1(N_m)\\\sum_{j=1}^M u_{j,m} / M \end{array} \right] \right)  \right) \ ,
\end{align}
\end{subequations}
where $f_1$ represents pumping, $\mathbf{f}_2$ the energy build-up and $f_{3,N}$ and $f_{3,E}$ the effect of coupling-out on inversion and pulse energy, respectively. 
Such explicit expressions provide an efficient tool for analyzing the resulting laser dynamics as well as designing real-time capable control and estimation algorithms.

\section{Simulation results}
\label{section:results}
In the following, we will show simulation results which demonstrate the interplay of stochastic and dynamic effects, agreement with first measurement results and the accuracy of the model simplifications proposed in the previous section.

\subsection{Pulse statistics and bifurcations}
\begin{figure}
\begin{center}
\includegraphics[width=0.68\textwidth]{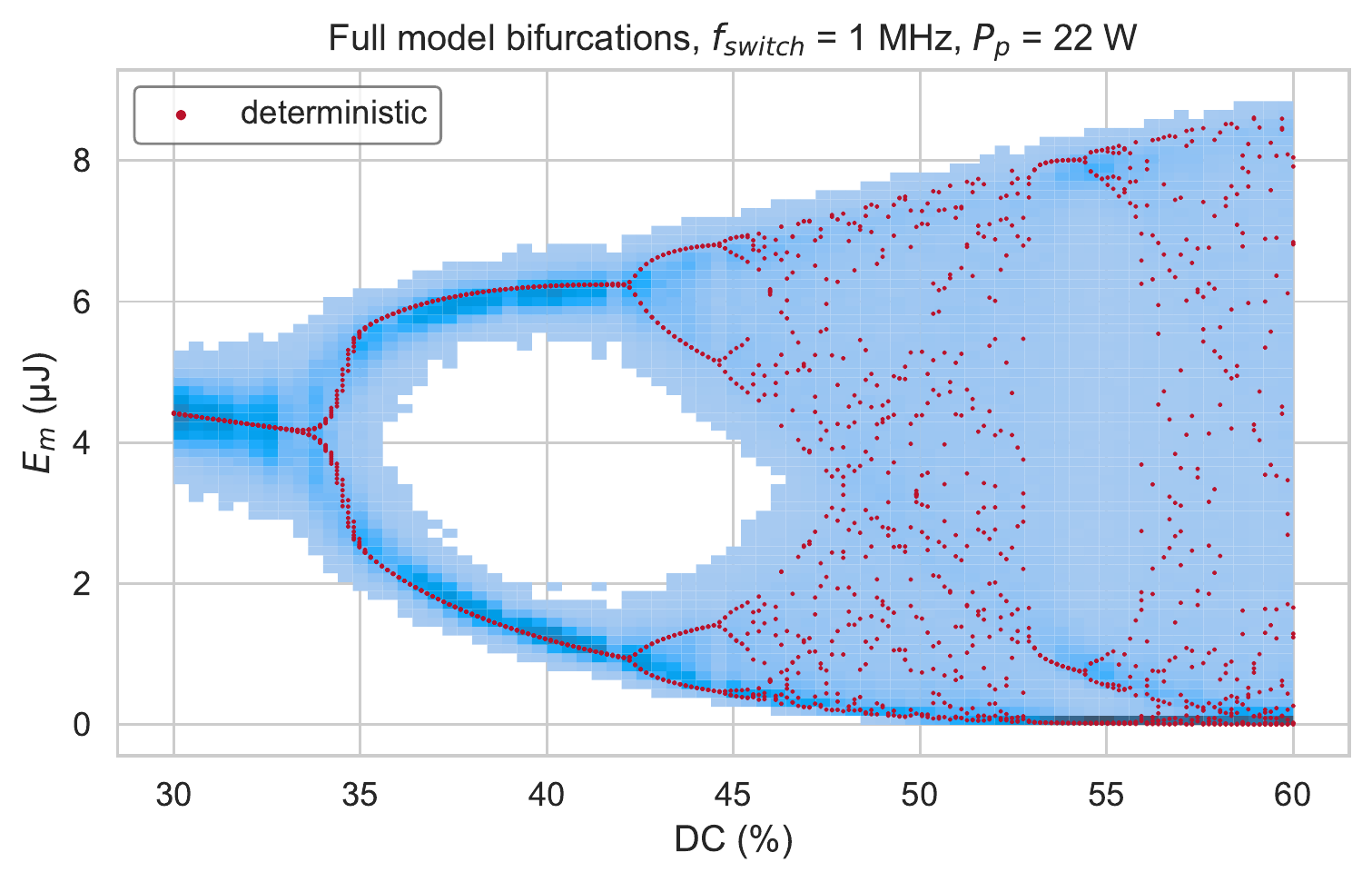}
\caption{Distribution of output pulse energies of the full stochastic model (\ref{equ:rate_equations}-\ref{equ:E_out}) and its deterministic counterpart for varying relative duty cycles (DC). As the DC increases, losses lower the mean output energy and period-doubling bifurcations occur. }
\label{fig:hist_dynamics}
\end{center}
\end{figure}

Apart from noise on the intracavity and output powers during a switching cycle, spontaneous emission also causes variations in the output energies even when the system is dynamically stable. To study the interplay of stochastic and dynamic effects in our general model discussed in Section \ref{section:general model} and given by (\ref{equ:rate_equations}-\ref{equ:E_out}), the duty cycle (DC) of the Pockels cell is varied at a fixed switching frequency ($f_{switch} =$ 1 MHz) and pump power ($P_p =$ 22 W). By plotting the output energy histograms and deterministic output pulses for comparison, a bifurcation diagram (Fig. \ref{fig:hist_dynamics}) is obtained which displays how period-doubling bifurcations and increasing losses due to a longer build-up affect the energy distribution. Since a cavity design with large bandwidth was assumed, we have a large number of contributing modes and hence expect roughly Poissonian output energy statistics, as was pointed out in \cite{Bose_Einstein_Poisson}, which is indeed the case, cf. Fig. \ref{fig:CTNM_accuracy}b).
\noindent On top of the stochastic variations, period-doubling bifurcations at relative duty cycles of around 34\% and 42\% can be observed. From the third period-doubling onwards, we get deterministic chaos with an intermittent window of a 3-limit cycle between 53 and 56 \%. The mean output energy diminishes at such build-up times since the amplification phase has no significant gain left and losses dominate instead.

\subsection{Validity of the CTNM}
\label{section:CTNM_accuracy}

\begin{figure}
\begin{center}
\includegraphics[width=0.68\textwidth]{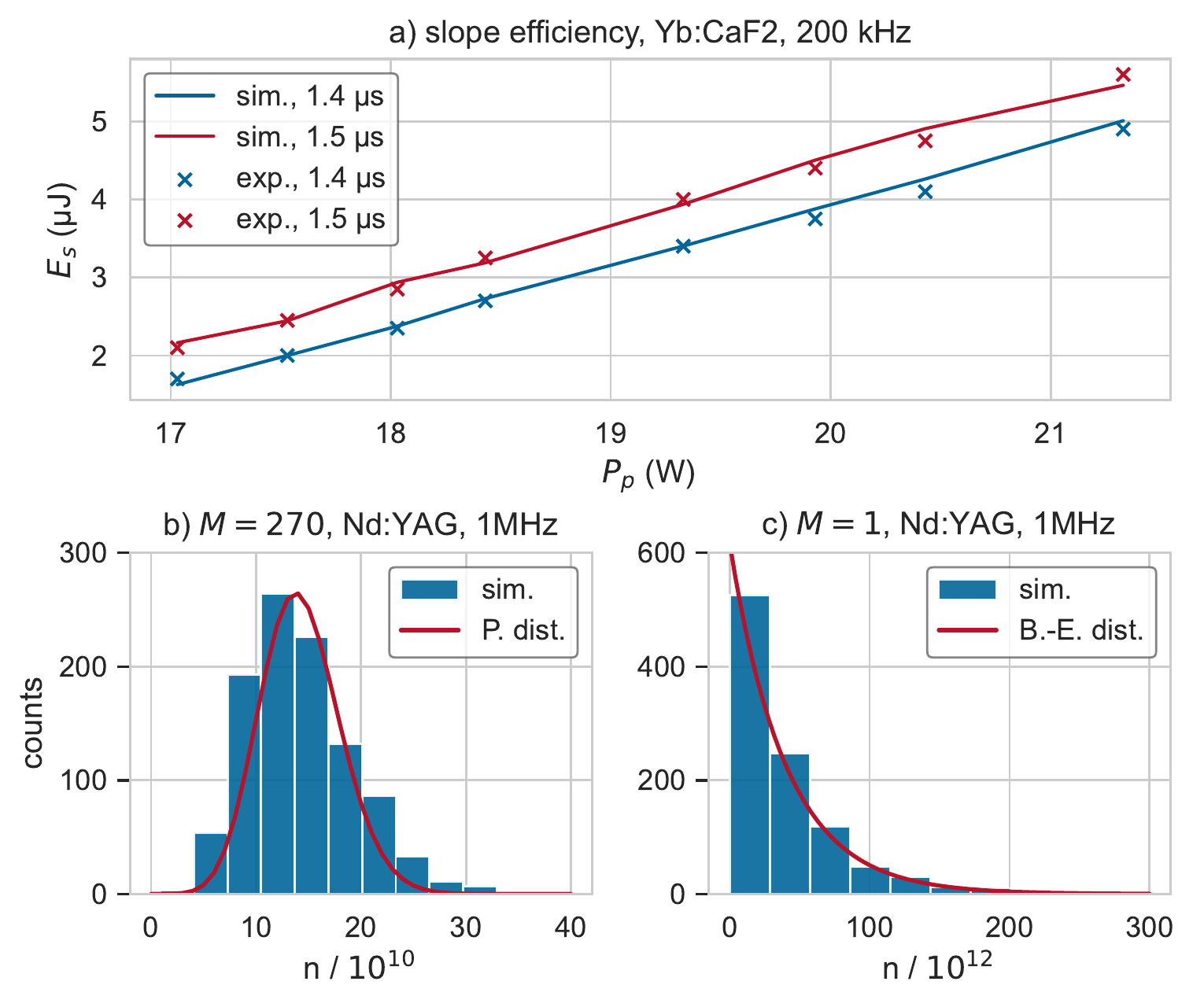}
\caption{Comparison of the output energies $E_m$ of (\ref{comp_red_rate},\ref{continuous_P_equation},\ref{equ:P_out_CTNM},\ref{equ:E_out}) to measured data and theoretical predictions \cite{Bose_Einstein_derivation}. a) shows the simulated steady-state energies $E_s$ and average energies from an experimental Yb:CaF2 laser. The bottom row compares the distribution of simulated normalized photon numbers (\ref{equ:normalized_photon_number}) obtained from output energies of an Nd:YAG laser to a) a Poisson ($M=270$) and b) a Bose-Einstein distribution with $M=1$. }
\label{fig:CTNM_accuracy}
\end{center}
\end{figure}

The continuous-time nonlinear model (CTNM) given by (\ref{comp_red_rate},\ref{continuous_P_equation}) together with (\ref{equ:P_out_CTNM},\ref{equ:E_out}) simplifies the spatially distributed full model (\ref{equ:rate_equations}-\ref{equ:E_out}) by averaging all processes within the cavity over half a round-trip such that ordinary differential equations remain to describe the whole system dynamics. We will show in Section \ref{section:model_simplification_comparison} (cf. Fig. \ref{fig:model_comparison}) that this introduces only small model errors as the full model and the CTNM agree well both in stable and unstable regimes of operation.
Fig. \ref{fig:CTNM_accuracy} demonstrates that the CTNM can reproduce experimental data in terms of dependence on $P_p$ and the DC as well as theoretical predictions regarding the photon statistics of output pulses. The measured average pulse energies shown in the top row stem from a double-pass diode-pumped Yb:CaF2 laser system at 200 kHz switching frequency and build-up times $t_{bu} = \textnormal{DC}/f_{switch}$ of 1.4 $\mu$s and 1.5 $\mu $s, respectively. The corresponding simulation data represent steady-state output energies $E_s$ of (\ref{equ:DTNM_general}) and were obtained using (\ref{comp_red_rate},\ref{continuous_P_equation},\ref{equ:P_out_CTNM},\ref{equ:E_out}) without stochastic effects, with technical parameters adapted to the experiments and physical parameters similar to those from the literature (for reference, see the supplemental document for the Nd:YAG laser simulation parameters). One can see that the model fits the measured data very well both for different pump powers and for different build-up times.
The bottom row of Fig. \ref{fig:CTNM_accuracy} shows the distribution of the normalized photon numbers
\begin{align} \label{equ:normalized_photon_number}
n := \frac{E_m \lambda}{M hc  }
\end{align}
obtained from stochastic pulse energies $E_m$ given by (\ref{equ:E_out}), similarly to how it was done in \cite{Bose_Einstein_Poisson} with experimental data. The histograms come from simulations of an Nd:YAG system at 1 MHz switching frequency and 20\% DC exhibiting stable pulse-to-pulse dynamics, where the cavity supports around 270 contributing longitudinal modes (bottom left) and a single mode (bottom right). For comparison, the red lines in the bottom plots of Fig. \ref{fig:CTNM_accuracy} represent a Poisson and a Bose-Einstein distribution and hence the theoretical predictions from \cite{Bose_Einstein_derivation}, respectively. It should also be noted here that the CTNM and the full model exhibit equivalent pulse energy statistics.

\subsection{Comparison of models}
\label{section:model_simplification_comparison}

\begin{figure}
\begin{center}
\includegraphics[width=0.68\textwidth]{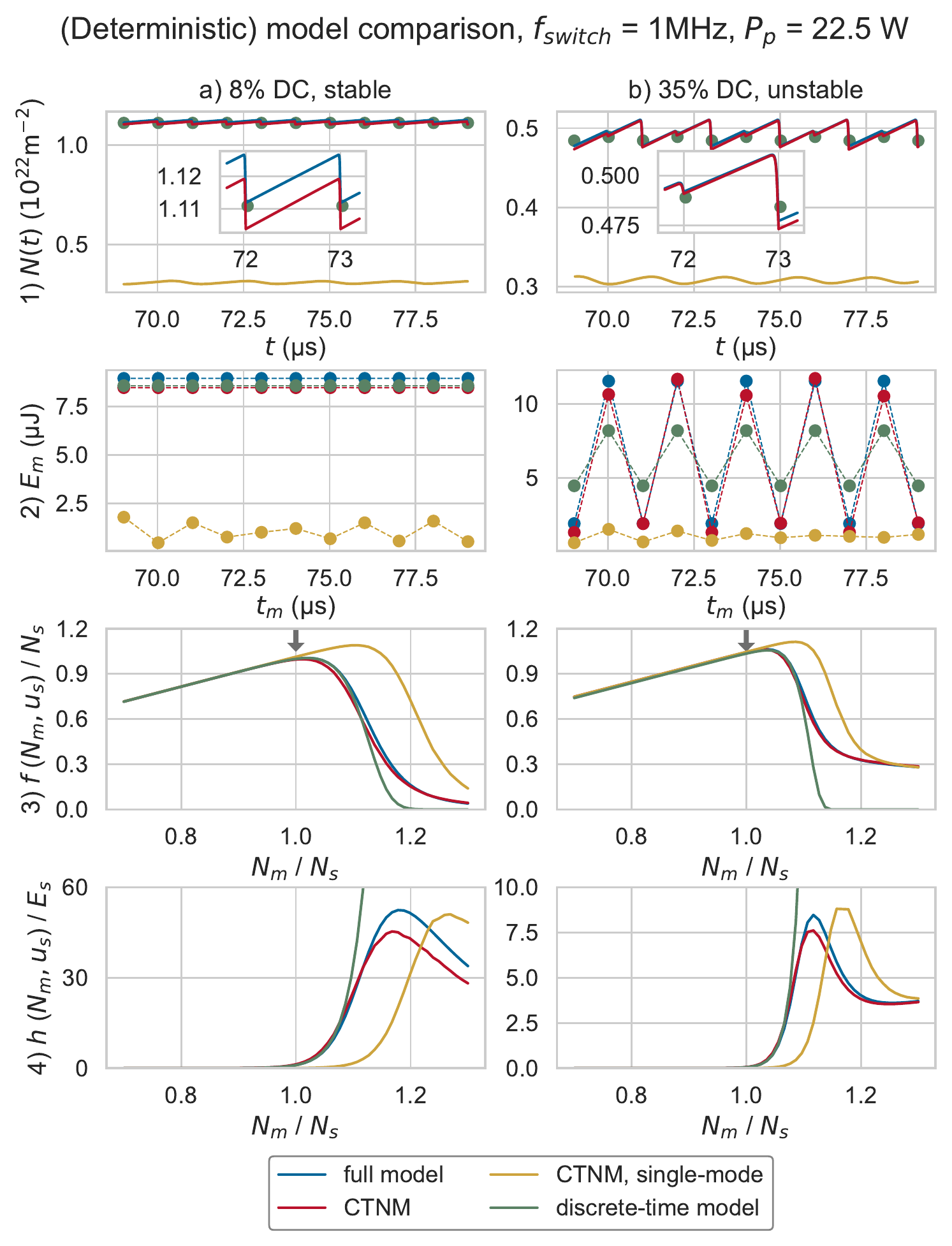}
\caption{Collection of simulation results of full model ((\ref{equ:rate_equations}-\ref{equ:E_out}), blue), continuous-time nonlinear model (CTNM, (\ref{comp_red_rate},\ref{continuous_P_equation},\ref{equ:P_out_CTNM},\ref{equ:E_out}), red), single-mode CTNM (yellow) and the discrete-time pulse-to-pulse model ((\ref{equ:DTNM}), green). The two columns a) and b) correspond to a dynamically stable and unstable scenario. The rows display 1) total population inversions, 2) pulse energiese and 3),4) normalised pulse-to-pulse dynamic maps for populations and output energies.}
\label{fig:model_comparison}
\end{center}
\end{figure}

Since we have shown in Section \ref{section:CTNM_accuracy} that the CTNM (\ref{comp_red_rate},\ref{continuous_P_equation},\ref{equ:P_out_CTNM},\ref{equ:E_out}) is able to reproduce the behavior of experimental results and theoretically predicted photon statistics of output pulses, we will now assess whether the CTNM differs significantly from the full model (\ref{equ:rate_equations}-\ref{equ:E_out}) and how further simplifications compromise the model's accuracy by analysing a collection of simulation results shown in Fig. \ref{fig:model_comparison}. One simplification lies in reducing the system to single-mode operation, neglecting all other cavity modes in (\ref{comp_red_rate},\ref{continuous_P_equation}) besides the dominating one (i.e. the mode subject to the highest gain). Another simplification of the CTNM is its approximate analytic solution, i.e. the discrete-time model (\ref{equ:DTNM}) of the pulse-to-pulse dynamics.
Fig. \ref{fig:model_comparison} contains two exemplary cases for a system at $f_{switch}$ = 1MHz repetition rate and a pump power $P_p$ = 22.5 W exhibiting stable (a: 8\% DC) and unstable (b: 35\% DC) pulse-to-pulse dynamics. They are simulated using the full model (\ref{equ:rate_equations}-\ref{equ:E_out}), the CTNM (\ref{comp_red_rate},\ref{continuous_P_equation},\ref{equ:P_out_CTNM},\ref{equ:E_out}), a single-mode version of the CTNM and the discrete-time pulse-to-pulse model (\ref{equ:DTNM}) discussed in Section \ref{section:DTNM}. Here, $N_s$, $u_s$ and $E_s$ stand for the steady-state population, uncontrolled intracavity power and pulse energy of the pulse-to-pulse dynamics (\ref{equ:DTNM_general}).
Row 3) of Fig. \ref{fig:model_comparison} demonstrates that increasing the DC value alters the shape of the pulse-to-pulse dynamics curve $f(N_m,u_s)/N_s$. Increasing one of the other two bifurcation parameters $f_{switch}$ or $P_p$ moves the steady state (position of the grey arrow) of the pulse-to-pulse dynamics further to the right of the dynamical map $f(N_m,u_s)$ (to higher $N_m$) such that eventually the absolute value of the slope becomes larger than one and the dynamics hence unstable. In Fig. \ref{fig:model_comparison}, we can see that the single-mode approximation (yellow) introduces major modeling errors as all photons are emitted into the channel with maximum gain. This in turn leads to a premature energy build-up with enormous depletion and power losses.
One can further see - most distinctly in the insets of 1a) and 1b) - that the population inversions of the full model and the CTNM differ only by about 1 percent. The discrete-time model matches these dynamics relatively well, but in the instability region of the pulse-to-pulse dynamics, the limit cycles for populations in 1b) as well as pulse energies in 2b) have smaller amplitudes. 
In terms of pulse energies $E_m$, row 2) again displays a good agreement between the CTNM and the full model. 
The normalised pulse-to-pulse dynamic maps in rows 3) and 4) of Fig. \ref{fig:model_comparison} for the full model, the CTNM and the discrete-time model stay in close proximity. However, it should be emphasized that the discrete-time model fits the dynamics well up to around 1.1 times the steady-state population $N_s$ in rows 3) and 4). At larger inversion, too much population is depleted in the discrete-time model, and the output map $h(N_m,u_s)$ overshoots significantly. This is due to the fact that losses can only be incorporated via operator splitting, as we showed in the supplemental document. In practice, the overshoot is not problematic at all as such high inversions are never reached if the cavity is being opened and closed periodically, but an analysis on a broad domain of inversion values is still useful for assessing the global dynamical behavior of the models.

\section{Conclusion and Outlook}
\label{section:conclusion}

In this work, a computationally efficient mathematical model for an actively Q-switched laser system including the effects of stochastic spontaneous emission is derived as a basis for the design of advanced control and estimation strategies. A spatially distributed multi-mode traveling-wave model, which is commonly used in the literature, serves as a starting point. Based on this model, the power transport equations were simplified by making physically feasible assumptions and the singular perturbation theory was applied to systematically reduce the number of relevant populations. Then all the processes within the cavity were averaged over half a cavity round-trip. As a result, a set of coupled, nonlinear ordinary differential equations were obtained, which describe the evolution of the inversion and each cavity laser mode at any point in time. Moreover, in order to separately study dynamic and stochastic effects on energy fluctuations, a discrete-time model of the pulse-to-pulse dynamics was derived by solving the differential equations in a semi-analytic way.
Simulation results show that the model simplification steps only introduce small errors in the range of sub-percent, except for the discrete-time model at large populations, which does not constitute an issue for the practical application. In particular, the systematic incorporation of multi-mode behavior and spatially distributed amplification, which distinguishes the spatially lumped continuous-time nonlinear model presented in this work from traditional point models known from the literature, provides superior modeling accuracy with respect to the full model at moderate computational costs. Thus, the continuous-time nonlinear model accurately describes the behavior of actively Q-switched lasers during a switching cycle, while the presented discrete-time model provides a simple relation between the initial populations and the corresponding output energies of two different cycles. The simulation results show that stochastic seeding leads to amplified spontaneous emission noise whose pulse statistics correspond to theoretical results from the literature, while the simulated slope efficiencies of output energies for various build-up times agree with first measurement results presented in the paper. Moreover, as it was expected from experimental observations and other papers from the literature, under certain conditions the nonlinear pulse-to-pulse dynamics exhibits period-doubling bifurcations and deterministic chaos.
The model presented in this paper aims at laying the foundation to systematically design control and estimation strategies to achieve robustly stable Q-switched lasers at high repetition rates. In particular, the pulse-to-pulse dynamics can be stabilised or otherwise modified via active feedback methods and prelasing strategies can be employed to suppress stochastic seeding due to spontaneous emission. \\

\appendix
\begin{center} \begin{LARGE}{A stochastic nonlinear model of the dynamics of actively Q-switched lasers: supplemental document} \end{LARGE} \end{center}

\section{Approximate solution of the power transport equations}
\label{section:power_transport_solution}

As mentioned in the paper, we will present here how the assumptions (A1.) and (A2.) lead to an approximate solution of the power transport equations (2) in the paper. In this section, we will omit wavelengths $\lambda_j$ in all function arguments for conciseness.\\
 Starting with (2), we introduce the moving frame by $t \rightarrow t^\mp := t \mp z/v$. As a consequence, the partial derivatives in $z$ transform as
 \begin{align*}
 \frac{\partial P^\pm (z,t)}{\partial z} \rightarrow  \frac{\partial P^\pm (z,t^\mp)}{\partial z}  \mp 1/v \frac{\partial P^\pm (z,t^\mp)}{\partial t^\mp}
 \end{align*}
by the chain rule (for the pump power analogously). The power transport equations then take the form
 \begin{subequations}
\begin{align}	
&  \parderiv{P_p(z,t^-)}{z}  = \sigma_{p} P_p(z,t^-) \left( N_5(z,t^-)-N_1(z,t^-) \right)-\alpha_p P_p(z,t^-)\label{powereqn1}\\
 & \pm \parderiv{P^\pm(z,t^\mp)}{z} = P^\pm(z,t^\mp)\Big( \sigma_A \left( N_4(z,t^\mp)-N_2(z,t^\mp) \right) \ +\label{powereqn2} \\
&  \sigma_B \left( N_4(z,t^\mp)-N_3(z,t^\mp) \right) \Big)  -\alpha P^\pm(z,t^\mp)  +\alpha_{RS} P^\mp(z,t^\mp) + \frac{2hc^2}{\lambda_j^3} \Delta \lambda_j \frac{\Delta\Omega}{4\pi} \zeta(z,t^\mp) \ .  \notag
\end{align}
\end{subequations}
We will now focus on forward-running signal beams (pump and backward-running beams work analogously). By assumption (A1.), (\ref{powereqn2}) can be solved analytically, and we get
\begin{align} \label{P_solution_before_fia}
&P^+(L) = \Bigg( \int_0^L \Bigg[ \textnormal{e}^{ -\int_0^z \Big( \sigma_A \left( N_4(z')-N_2(z') \right)  + \sigma_B \left( N_4(z')-N_3(z') \right) -\alpha \Big) \textnormal{d}z'   } \times \\
&\times \Big( \alpha_{RS} P^-(L) + \frac{2hc^2}{\lambda_j^3} \Delta \lambda_j \frac{\Delta\Omega}{4\pi} \zeta(z) \Big) \Bigg] \ \textnormal{d} z + P^+(0) \Bigg)   \textnormal{e}^{ \int_0^L \Big( \sigma_A \left( N_4(z)-N_2(z) \right)  + \sigma_B \left( N_4(z)-N_3(z) \right) -\alpha \Big) \textnormal{d}z   } \ .   \notag
\end{align} 
We see that the exponential within the integral cancels a part of the outer exponential. This means that spontaneously emitted or backscattered photons experience a gain which depends on where they are emitted. By virtue of assumption (A2.), photons joining the $P^+$beam do so at the end of the active medium ($z=L$), and hence, 
\begin{align} \label{fia}
\textnormal{e}^{ -\int_0^z \Big( ... \Big) \textnormal{d}z' } \approx \textnormal{e}^{ -\int_0^L \Big( ... \Big) \textnormal{d}z' } \ .
\end{align}
Thus, the exponentials exactly cancel, and writing out time arguments again, we can directly evaluate all integrals since we have defined the total populations $N_i^{tot}(t)$ as integrals over $N_i(z,t)$ and the photon count $ \mathfrak{C}(t)$ as integral over $\zeta(z,t) $. As written in the paper, the wavelength spacing is chosen correspondingly to each two frequencies being separated by an inverse cavity round-trip time. Hence, we obtain
\begin{align} \label{wavelength_spacing}
\Delta \lambda_j = \frac{\lambda_j^2}{c} \Delta \nu =  \frac{\lambda_j^2}{c} / t_{RT} \ ,
\end{align}
which, if we insert it into (\ref{P_solution_before_fia}) together with (\ref{fia}), yields the solutions given by (12a,12b) in the paper.

\section{Derivation of the total rate equations}

We will now derive the total rate equations given by (13a-13e) in the paper by using assumptions (A1.)-(A4.). For demonstrational purposes, we will show the derivation of $\deriv{N_3^{tot}(t)}{t}$, but the concept applies to all equations equivalently and the computations are analogous. \\
First, we write the rate equation (1c) as
\begin{align}
	\label{popeqn3}\begin{split}& \parderiv{N_3(z,t)}{t} = -\gamma_{31} N_3(z,t) + \gamma_{13} N_1(z,t) + \gamma_{43} N_4(z,t) \\
	& +  \sum_{j=1}^M \frac{ \lambda_j}{h c A_s} \left(P^{+}(z,t,\lambda_j) + P^{-}(z,t,\lambda_j)\right)\Big(\sigma_B(\lambda_j) \left(N_4(z,t) - N_3(z,t)\right)  \Big) \ . \end{split}
\end{align}
For the second row of (\ref{popeqn3}), we implement assumption (A4.) by employing the lineshape separator function $g_B(\lambda)$. For the rate equation (1d), this would be $g_A(\lambda)$, and for (1b), it would be $g_A(\lambda)+g_B(\lambda) \equiv 1$. This means that in (\ref{popeqn3}), we have
\begin{align} \label{Gamma_definition}
 & \left(P^{+}(z,t,\lambda_j) + P^{-}(z,t,\lambda_j)\right)\Big(\sigma_B(\lambda_j) \left(N_4(z,t) - N_3(z,t)\right)  \Big)  = g_B(\lambda_j) \Gamma(z,t,\lambda_j) := \\
& = g_B(\lambda_j)  \left(P^{+}(z,t,\lambda_j) + P^{-}(z,t,\lambda_j)\right)\Big(\sigma_A(\lambda_j) \left(N_4(z,t) - N_2(z,t)\right) + \sigma_B(\lambda_j) \left(N_4(z,t) - N_3(z,t)\right)  \Big) \ . \notag
\end{align}
If we now look at the power transport equation in a moving frame, i.e. (\ref{powereqn2}), with (\ref{wavelength_spacing}) already applied, we observe that
\begin{align} \label{Gamma_with_powereqn}
\Gamma(z,t,\lambda_j) &= \parderiv{P^+(z,t,\lambda_j)}{z} -  \parderiv{P^-(z,t,\lambda_j)}{z} + \\
 & \left( \alpha(\lambda_j)  - \alpha_{RS}\right)  \left(P^{+}(z,t,\lambda_j) + P^{-}(z,t,\lambda_j)\right) - 4 \frac{hc}{\lambda_j t_{RT}} \frac{\Delta \Omega}{4\pi} \zeta(z,t,\lambda_j) \ . \notag
\end{align}
Inserting (\ref{Gamma_definition}) into (\ref{popeqn3}), which yields
\begin{align*}
& \parderiv{N_3(z,t)}{t} = -\gamma_{31} N_3(z,t) + \gamma_{13} N_1(z,t) + \gamma_{43} N_4(z,t) +  \sum_{j=1}^M g_B(\lambda_j) \frac{ \lambda_j}{h c A_s} \Gamma(z,t,\lambda_j) \ ,
\end{align*}
and integrating both sides over $z$, the integrals of the time derivative and relaxation terms can be performed immediately due to the definition of $N_i^{tot}(t)$ in (11) of the paper. For the depletion term $\Gamma(z,t,\lambda_j)$, the integral of (\ref{Gamma_with_powereqn}) reads
\begin{align} \label{Gamma_integral} \begin{split}
\int_0^L \Gamma(z,t,\lambda_j) \ \textnormal{d}z &= P^+(L,t,\lambda_j) - P^+(0,t,\lambda_j) + P^-(0,t,\lambda_j) - P^-(L,t,\lambda_j) \ + \\
& \int_0^L \left( \alpha(\lambda_j)  - \alpha_{RS}\right)  \left(P^{+}(z,t,\lambda_j) + P^{-}(z,t,\lambda_j)\right) \ \textnormal{d}z - 4 \frac{hc}{\lambda_j t_{RT}} \frac{\Delta \Omega}{4\pi} \mathfrak{C}(t,\lambda_j) \ , \end{split}
\end{align}
where we used the definition of the stochastic term provided in (10) of the paper. For $ P^+(L,t,\lambda_j)$ and $P^-(0,t,\lambda_j)$ in (\ref{Gamma_integral}), we insert the power transport solution given by (12b) in the paper and derived in Section \ref{section:power_transport_solution}. The spontaneous emission terms then cancel, as do the $\alpha_{RS}$ terms due to assumption (A1.). The integral of the term in (\ref{Gamma_integral}) which involves $\alpha(\lambda_j)$ can be directly computed because of assumption (A3.), and we obtain
\begin{align*}
\int_0^L  \alpha(\lambda_j)  \left(P^{+}(z,t,\lambda_j) + P^{-}(z,t,\lambda_j)\right) \textnormal{d}z =  \alpha(\lambda_j) L \left(P^{+}(0,t,\lambda_j) + P^{-}(L,t,\lambda_j)\right) \ ,
\end{align*}
which implies 
\begin{align*}
& \int_0^L \Gamma(z,t,\lambda_j) \ \textnormal{d}z =  \left(P^+(0,t,\lambda_j)+P^-(L,t,\lambda_j) \right) \times  \\
 & \times \bigg( \exp \Big( \sigma_A(\lambda_j) \left(N_4^{tot}(t) - N_2^{tot}(t)\right) + \sigma_B(\lambda_j) \left(N_4^{tot}(t) - N_3^{tot}(t)\right) -\alpha(\lambda_j)L \Big) -1 + \alpha(\lambda_j)L \bigg) \ .
\end{align*}
Finally, the resulting differential equation yields (13c) if we make the substitution
\begin{align*}
G(t,\lambda_j) := \frac{\lambda_j}{hcA_s} \int_0^L \Gamma(z,t,\lambda_j) \ \textnormal{d}z \ .
\end{align*}

\section{State reduction via slow-fast dynamics}

In the paper, we stated that the reduced total rate equation (17) can be derived as a quasi-stationary model from (13) by applying singular perturbation theory. More concretely, the limit of infinitely fast relaxation within the upper and lower state manifolds which is expressed by (14) leads to a slow and a fast subsystem of (13). We will now pursue this path in detail, derive (17) and assess the approximation order of the reduced model via Tikhonov's theorem (Theorem 11.4 in \cite{Khalil}). \\
In this section, we introduce the Boltzmann distribution thermalization factors $B_{ij} := \exp\left(\frac{-\Delta E_{ij}}{k_BT}\right)$ and use the abbreviations
\begin{align*}
 N_i &:= N_i^{tot}(t)\ \forall i \in \{ 1,2,3,4,5 \} \\
\gamma &:= \gamma_{42} + \gamma_{43} \\
b &:= 1/(1+B_{45}) \\
B&:= 1/(1+B_{12}+B_{13}) \\
H(x)&:= \frac{\lambda_{p}}{h c A_p} \bigg(1 - \exp \Big( \sigma_{p} x -\alpha_pL \Big) -\alpha_pL \bigg) P_p(t) \\
G_j(x,y) &:= \frac{\lambda_j}{h c A_s} \left(P^+(0,t,\lambda_j)+P^-(L,t,\lambda_j) \right) \times  \\
 & \times \bigg( \exp \Big( \sigma_A(\lambda_j) x + \sigma_B(\lambda_j) y -\alpha(\lambda_j)L \Big) -1 + \alpha(\lambda_j)L \bigg) \\
 g_A(j) &:= g_A(\lambda_j)\\
 g_B(j) &:= g_B(\lambda_j) \\
 \sum_{j=1}^M &:= \sum_j \ .
\end{align*}
The total rate equations (13) then read as
\begin{subequations}\label{total_rate_start}
\begin{align} 
\deriv{N_5}{t} &= -\gamma_{54} N_5 + \gamma_{45} N_4 + H(N_5-N_1) \\
\deriv{N_4}{t} &= -(\gamma_{45}+\gamma_{43}+\gamma_{42}) N_4 + \gamma_{54} N_5 - \sum_j G_j(N_4-N_2,N_4-N_3) \\
\deriv{N_3}{t} &= -\gamma_{31} N_3 + \gamma_{13} N_1 + \gamma_{43} N_4 + \sum_j g_B(j) G_j(N_4-N_2,N_4-N_3) \\
\deriv{N_2}{t} &= -\gamma_{21} N_2 + \gamma_{12} N_1 + \gamma_{42} N_4 + \sum_j g_A(j) G_j(N_4-N_2,N_4-N_3) 
\end{align}
\end{subequations}
with $N_1 = LN_{dop} - N_5-N_4-N_3-N_2$. \\
As a next step, we employ a linear state transformation
\begin{subequations}\label{N_to_y_trafo}
\begin{align}
y_5 &= N_5 - B_{45}N_4 \\
y_4 &= N_5 + N_4 \\
y_3 &= N_3 - B_{13}N_1 \\ 
y_2 &= N_2 - B_{12}N_1 \\ 
y_1 &= N_1+N_2+N_3 \ ,
 \end{align}
 \end{subequations}
where we have $y_4+y_1 = LN_{dop}$. 
Inversion of the transformation yields
\begin{subequations}\label{y_to_N_trafo}
\begin{align}
N_5 &= (1-b)y_4+by_5 \\
N_4 &= b(y_4-y_5) \\
N_3 &= (1+B_{12})By_3 - B_{13}By_2 + B_{13}By_1\\ 
N_2 &= -B_{12}By_3+(1+B_{13})By_2 + B_{12}By_1\\ 
N_1 &= B(y_1-y_2-y_3) \ .
 \end{align}
 \end{subequations}
By differentiation of (\ref{N_to_y_trafo}) and insertion of (\ref{total_rate_start}), one obtains
\begin{subequations}\label{y_rate_equation_before_insertion}
\begin{align}
\begin{split} \deriv{y_5}{t} &= -(1+B_{45}) \gamma_{54} N_5 + \gamma_{45} N_4 + B_{45} (\gamma_{45}+\gamma_{43}+\gamma_{42})N_4 \\
& + H(N_5-N_1) + B_{45}\sum_j G_j(N_4-N_2,N_4-N_3) \end{split} \\ 
\deriv{y_4}{t} &= -(\gamma_{43}+\gamma_{42}) N_4 +  H(N_5-N_1) - \sum_j G_j(N_4-N_2,N_4-N_3)  \\
\begin{split} \deriv{y_3}{t} &= -(1+B_{13})\gamma_{31}N_3 + \gamma_{43}N_4 - B_{13}\gamma_{21}N_2 + (1+B_{13})\gamma_{13}N_1 + B_{13}\gamma_{12}N_1  \\
& + B_{13} H(N_5-N_1) + \sum_j g_B(j) G_j(N_4-N_2,N_4-N_3) \end{split}  \\
\begin{split} \deriv{y_2}{t} &= -(1+B_{12})\gamma_{21}N_2 + \gamma_{42}N_4 - B_{12}\gamma_{31}N_3 + (1+B_{12})\gamma_{12}N_1 + B_{12}\gamma_{13}N_1 \\
& + B_{12} H(N_5-N_1) + \sum_j g_A(j) G_j(N_4-N_2,N_4-N_3) \ . \end{split} 
\end{align}
\end{subequations}
After inserting (\ref{y_to_N_trafo}) into (\ref{y_rate_equation_before_insertion}), we set up the slow-fast dynamics by assuming thermal equilibrium, i.e.
\begin{align*}
\gamma_{45} = B_{45}\gamma_{54} , \ \gamma_{13} = B_{13}\gamma_{31} , \ \gamma_{12} = B_{12}\gamma_{21} \ , 
\end{align*}
as well as introducing the ratios
\begin{align*}
\gamma_{31} = a_{31} \gamma_{54}, \ \gamma_{21} = a_{21} \gamma_{54}, \  a_{31} > 0, \  a_{21} > 0 \ .
\end{align*}
If we now define the singular perturbation parameter $\epsilon := 1/\gamma_{54}$, we get
\begin{subequations} \label{almost_slow_fast}
\begin{align}
\begin{split} \deriv{y_5}{t} &= -\frac{1}{\epsilon} \bigg( (1+B_{45}) y_5 \bigg) + bB_{45} \gamma (y_4-y_5) \\
& + H(N_5-N_1) + B_{45} \sum_j G_j(N_4-N_2,N_4-N_3) \label{almost_slow_fast_5} \end{split}  \\
\deriv{y_4}{t} &= -b \gamma (y_4-y_5) + H(N_5-N_1) - \sum_j G_j(N_4-N_2,N_4-N_3) \\
\begin{split} \deriv{y_3}{t} &= -\frac{1}{\epsilon} \bigg( B_{13} a_{21} y_2 + (1+B_{13})a_{31}y_3 \bigg) + b \gamma_{43} (y_4-y_5) \label{almost_slow_fast_3}  \\
& + B_{13} H(N_5-N_1) +  \sum_j g_B(j) G_j(N_4-N_2,N_4-N_3) \end{split}  \\
\begin{split} \deriv{y_2}{t} &= -\frac{1}{\epsilon} \bigg( B_{12} a_{31} y_3 + (1+B_{12})a_{21}y_2 \bigg) + b \gamma_{42} (y_4-y_5) \\
& + B_{12} H(N_5-N_1) +  \sum_j g_A(j) G_j(N_4-N_2,N_4-N_3) \ , \label{almost_slow_fast_2}  \end{split} 
\end{align}
\end{subequations}
where we have not transformed the arguments of $H$ and $G_j$ for brevity. Multiplying (\ref{almost_slow_fast_5},\ref{almost_slow_fast_3},\ref{almost_slow_fast_2}) by $\epsilon$ and defining the slow state variable 
\begin{align*}
x := y_4
\end{align*}
as well as the fast state variables
\begin{align*}
\mathbf{z} := \left[ \begin{array}{c}
y_5 \\
y_3 \\
y_2
\end{array} \right] \ ,
\end{align*}
the system (\ref{almost_slow_fast}) takes the singular perturbation standard form
\begin{subequations} \label{singular_standard_form}
\begin{align}
\deriv{x}{t} &= f_1(t,x,\mathbf{z},\epsilon) \\
\epsilon \deriv{\mathbf{z}}{t} &= \mathbf{f}_2 (t,x,\mathbf{z},\epsilon) \ ,
\end{align}
\end{subequations}
with the dynamical maps
\begin{subequations} \label{singular_dyn_maps}
\begin{align}
f_1(t,x,\mathbf{z},\epsilon)  &= -b \gamma (y_4-y_5) + H(N_5-N_1) - \sum_j G_j(N_4-N_2,N_4-N_3)  \\
\mathbf{f}_2(t,x,\mathbf{z},\epsilon)   &= \left[ \begin{array}{c}
-(1+B_{45}) y_5  \\
+ \epsilon \bigg( bB_{45} \gamma (y_4-y_5) + H(N_5-N_1) + B_{45} \sum_j G_j(N_4-N_2,N_4-N_3) \bigg) \\
\\
- ( B_{13} a_{21} y_2 + (1+B_{13})a_{31}y_3 ) \\
+ \epsilon \bigg( b \gamma_{43} (y_4-y_5) + B_{13} H(N_5-N_1) +  \sum_j g_B(j) G_j(N_4-N_2,N_4-N_3) \bigg) \\
\\
-( B_{12} a_{31} y_3 + (1+B_{12})a_{21}y_2 )\\
+ \epsilon \bigg( b \gamma_{42} (y_4-y_5) + B_{12} H(N_5-N_1) +  \sum_j g_A(j) G_j(N_4-N_2,N_4-N_3) \bigg)
\end{array} \right] \ . \label{singular_dyn_f_2}
\end{align}
\end{subequations}
The quasi-stationary model is derived from (\ref{singular_standard_form}) via the limit $\epsilon \rightarrow 0$ and reads as
\begin{align*}
\deriv{x_r}{t} = f_1(t,x_r,\mathbf{z}_r,0) \ ,
\end{align*}
with $\mathbf{z}_r = \mathbf{g}(t,x_r)$ as the solution of the algebraic equations $\mathbf{f}_2(t,x_r,\mathbf{z}_r,0) = \mathbf{0}$. This yields
\begin{align*}
\mathbf{z}_r = \mathbf{g}(t,x_r) = \mathbf{0} 
\end{align*}
and hence the quasi-stationary model reads as
\begin{align} \label{quasi-stationary_model}
\deriv{x_r}{t} = f_1(t,x_r,\mathbf{0},0) \ .
\end{align}
According to Tikhonov's theorem (see Theorem 11.4 in \cite{Khalil}), the so-called boundary layer model
\begin{align}
\deriv{\mathbf{z}_s}{\tau} &=  \mathbf{f}_2 (t,x_s,\mathbf{z}_s,0) \ ,
\end{align}
with 
\begin{align*}
 \mathbf{f}_2 (t,x,\mathbf{z},0) = \left[
 \begin{array}{c}
 -(1+B_{45}) y_5  \\
 - ( B_{13} a_{21} y_2 + (1+B_{13})a_{31}y_3 ) \\
 -( B_{12} a_{31} y_3 + (1+B_{12})a_{21}y_2 )
 \end{array}
 \right] \ .
\end{align*}
in the fast time scale $\tau$ must be uniformly exponentially stable in $t$ and $x_s$, which is the case because
\begin{align*}
\lVert \mathbf{z}_s(\tau) \rVert \leq \lVert \mathbf{z}_s(0) \rVert \exp\left(- \textnormal{min} \big\{ 1+B_{45} , (1+B_{13})a_{31} , (1+B_{12})a_{21} \big\} \tau \right) \ .
\end{align*}
Applying Tikhonov's theorem further proves that there exist finite values of $\epsilon$ up to which the reduced system (\ref{quasi-stationary_model}) is of approximation order $\mathcal{O}(\epsilon)$ with respect to (\ref{singular_standard_form}). For more details on the singular perturbation theory, the reader is referred to, e.g., \cite{Khalil}. \\
Finally, the slow dynamics (\ref{quasi-stationary_model}) is transformed back to $N_i$ via (\ref{y_to_N_trafo}) and by employing $y_5 = y_3 = y_2 = 0$. This yields
\begin{align*}
N_4 &= by_4 \\
N_5 &= B_{45} N_4 \\
N_1 &= By_1 = B(LN_{dop}-N_4/b) \\
N_2 &= B_{12} N_1 \\
N_3 &= B_{13} N_1 \ ,
\end{align*}
which, keeping in mind that $y_4 = x$ and defining $N := N_4$, endows us with the reduced total rate equation 
\begin{align*}
&\deriv{ N}{t} =   - b\gamma N -  b\sum_{j=1}^M\frac{\lambda_j}{h c A_s} \Bigg( \exp\bigg( \sigma_A(\lambda_j)\left( N-BB_{12} (LN_{dop}-(1+B_{45})N) \right) + \sigma_B(\lambda_j) \times\\
&\times \left( N-BB_{13} (LN_{dop}-(1+B_{45})N) \right) -\alpha(\lambda_j)L  \bigg) -1 + \alpha(\lambda_j)L \Bigg)  \left(P^+(0,t,\lambda_j)+P^-(L,t,\lambda_j) \right) \\
&+ \frac{b\lambda_p}{hcA_p} \Bigg( 1-\exp\bigg( \sigma_p \left( B_{45}N -B(LN_{dop}-(1+B_{45})N) \right) -\alpha_pL  \bigg) -\alpha_pL  \Bigg) P_p(t) \ .
\end{align*}
One final approximation lies in setting $B_{12} = B_{13} = 0$ (large energy gaps $\Delta E_{12}$ and $\Delta E_{13}$ such that upwards relaxation is impossible), which entails $B=1$ and after which we arrive at (17) in the paper.

\section{Continuous-time nonlinear model}

We will now show how (20) can be obtained from (18) and (19) in the paper. \\
First, we write (18) for forward- and backward- running powers separately, i.e.
\begin{subequations}
\begin{align} \begin{split} \label{P_roundtrip_1}
&P^+\left(0,t_{k+1},\lambda_j\right) =\sqrt{ \eta(\lambda_j) R(t_k)} \exp\left(\sigma(\lambda_j) N(t_k)-\alpha(\lambda_j)L\right) \times \\
& \times \left( P^-\left(L,t_k,\lambda_j\right) +   2 \frac{hc}{\lambda_j t_{RT}} \frac{\Delta\Omega}{4\pi} \mathfrak{C}(t_k,\lambda_j) +\alpha_{RS} LP^-\left(L,t_k,\lambda_j\right) \right)  \end{split} \\
\begin{split} \label{P_roundtrip_2} &P^-\left(L,t_{k+1},\lambda_j\right) =\sqrt{ \eta(\lambda_j) R(t_k)} \exp\left(\sigma(\lambda_j) N(t_k)-\alpha(\lambda_j)L\right) \times \\
& \times \left( P^+\left(0,t_k,\lambda_j\right) +   2 \frac{hc}{\lambda_j t_{RT}} \frac{\Delta\Omega}{4\pi} \mathfrak{C}(t_k,\lambda_j) +\alpha_{RS} LP^+\left(0,t_k,\lambda_j\right) \right) \ . \end{split} 
\end{align}
\end{subequations}
Taking the sum of (\ref{P_roundtrip_1}) and (\ref{P_roundtrip_2}) and using the definition (16) of $P_j(t)$, we obtain
\begin{align}\begin{split} \label{P_j_roundtrip}
P_j(t_{k+1}) &= \sqrt{ \eta(\lambda_j) R(t_k)} \exp\left(\sigma(\lambda_j) N(t_k)-\alpha(\lambda_j)L\right) \times \\
& \times \left( P_j(t_k)+   4 \frac{hc}{\lambda_j t_{RT}} \frac{\Delta\Omega}{4\pi} \mathfrak{C}(t_k,\lambda_j) +\alpha_{RS} L P_j(t_k) \right) \ ,\end{split}
\end{align} 
and thus,
\begin{align} \label{P_j_logarithm_difference}
& \ln(P_j(t_{k+1})) - \ln(P_j(t_{k})) = \ln\left( \frac{P_j(t_{k+1})}{P_j(t_{k})} \right) \\
& = \ln\Bigg(  \sqrt{ \eta(\lambda_j) R(t_k)} \exp\left(\sigma(\lambda_j) N(t_k)-\alpha(\lambda_j)L\right)  \left( 1 +   4 \frac{hc}{\lambda_j t_{RT}} \frac{\Delta\Omega}{4\pi} \mathfrak{C}(t_k,\lambda_j) / P_j(t_k) +\alpha_{RS} L \right)  \Bigg) \notag
\end{align}
by virtue of (\ref{P_j_roundtrip}). The argument of the logarithm in (\ref{P_j_logarithm_difference}) consits of three factors which can be separated via $\ln(abc) = \ln(a)+\ln(b)+\ln(c)$. Multiplication of (\ref{P_j_logarithm_difference}) by $\frac{P_j}{t_{RT}/2}$ yields the first two terms of (20). The third and fourth terms come from the approximation that spontaneous emission and backscattering are small effects, i.e.
\begin{align*}
\frac{2P_j}{t_{RT}}\ln\left( 1 +   4 \frac{hc}{\lambda_j t_{RT}} \frac{\Delta\Omega}{4\pi} \mathfrak{C}(t_k,\lambda_j) / P_j +\alpha_{RS} L \right) \approx \frac{2}{t_{RT}}\left( 4 \frac{hc}{\lambda_j t_{RT}} \frac{\Delta\Omega}{4\pi} \mathfrak{C}(t_k,\lambda_j)  +\alpha_{RS} L P_j \right) \ ,
\end{align*}
by a Taylor approximation, which completes the derivation of (20).\\
In (21) of the paper, we give an expression for the output power as a function of the intracavity power under the approximation of disregarding time delays inside the cavity as well as Rayleigh backscattering. These approximations must be made to avoid an infinite regression in time and obtain a closed-form expression. \\
The given formula can be derived by firstly stating that forward- and backward-running powers must equal the intracavity power and secondly applying the boundary conditions (3,4) as well as the power transport solutions (12) given in the paper. In total, this means that we first solve the system of equations that is
\begin{align*}
 P_j(t) &= P^+(0,t,\lambda_j) + P^-(L,t,\lambda_j)  \\
P^-(L,t,\lambda_j) &=  \left( \exp\left( \sigma(\lambda_j)N(t) -\alpha(\lambda_j)L \right) P^+(0,t,\lambda_j) +  2\frac{hc}{\lambda_j t_{RT}} \frac{\Delta \Omega}{4\pi} \mathfrak{C}(t,\lambda_j) \right) \sqrt{\eta(\lambda_j)} R(t)  \ .
\end{align*} 
Note that by (4a), $R(t)$ appears here instead of its square root, which was applied to both forward- and backward-running beams, e.g. in (\ref{P_roundtrip_1}) and (\ref{P_roundtrip_2}), as an approximation to derive (\ref{P_j_roundtrip}). Solving for $P^-(L,t,\lambda_j)$, we get the output power from the backward-running power by first applying (4a), i.e.
\begin{align}
P^-(z_{PC},t,\lambda_j) = P^-(L,t,\lambda_j) / R(t) \ ,
\end{align}
and then (4b), which together reads as
\begin{align*}
P_{out}(t,\lambda_j) = \frac{1-R(t)}{R(t)} P^-(L,t,\lambda_j) \ .
\end{align*}
This yields the result (21) given in the paper.

\section{Compound Poisson process for spontaneous emission}
In order to model spontaneous emission occurring in the active medium, it is intuitively clear that some stochastic jump process will be necessary. In Section 2.3 of the paper, we state that it has been derived from a highly simplified model of amplifier chains that the number of emissions into a single mode ($M=1$) is Bose-Einstein distributed with a mean number $\mu$ of emitted photons. The variance of this distribution is $\mu^2 + \mu$. \\
A jump process which can be designed to fit these two moments is the compound Poisson process given by
\begin{align}\label{compound_Poisson_process}
X_t := \sum_{n=1}^{N_t} Y_n \ ,
\end{align}
where $N_t$ is a Poisson process with exponentially distributed waiting times. $Y_n$ are identically distributed random numbers independent of $N_t$. The choice $Y_n \equiv 1$ yields a Poisson process.\\
In view of the continuous-time nonlinear model from Section 3.3, we propose average waiting times of $t_{RT}/2$ where an average of $\sigma(\lambda_j)N(t)$ (see Section 2.3) photons is emitted per longitudinal mode $j$. This information can be linked to (\ref{compound_Poisson_process}) using Wald's formula such that \cite{Compound_Poisson} 
\begin{align}\label{Wald_equation}
\mathbb{E}(X_t) = r t \mathbb{E}(Y_1) \ ,
\end{align}
which suggests the rate parameter $r = 2/t_{RT}$ and $\mathbb{E}(Y_1) = \sigma(\lambda_j)N(t)$. As mentioned above, the Bose-Einstein distribution fixes the variance of emitted photons after $t_{RT}/2$ to be $(\sigma(\lambda_j)N(t))^2 +  \sigma(\lambda_j)N(t) $. For (\ref{compound_Poisson_process}), applying the Blackwell-Girshick equation yields \cite{Compound_Poisson}
\begin{align*}
\textnormal{Var}(X_t) = rt (\mathbb{E}(Y_1)^2 + \textnormal{Var}(Y_1)) \ .
\end{align*}
As a consequence, consistency with the Bose-Einstein distribution and $\mathbb{E}(Y_1) = \sigma(\lambda_j)N(t)$ requires
\begin{align*}
 \textnormal{Var}(Y_1) = \sigma(\lambda_j)N(t) \ .
\end{align*}
Therefore, $Y_n$ must be random numbers with both mean and variance equal to $\sigma(\lambda_j)N(t)$. This is satisfied by the Poisson distribution. We conclude that the stochastic process we use to model spontaneous emission is a compound Poisson process $X_t$ with rate parameter $2/t_{RT}$ and $Y_n$ drawn from a Poisson distribution with intensity parameter $\sigma(\lambda_j)N(t)$. \\
The noise term $\mathfrak{C}(t,\lambda_j)$ is finally obtained from the (rigorously non-existing) derivative
\begin{align*}
\mathfrak{C}(t,\lambda_j) = \frac{t_{RT}}{2} \deriv{X_t}{t} \ ,
\end{align*}
with the factor $\frac{t_{RT}}{2}$ for agreement with (20).

\section{Derivation of the discrete-time model}

We stated in the paper that the discrete-time pulse-to-pulse dynamics can be approximately expressed as
\begin{subequations}
\begin{align}\label{discrete_model_1}
N_{m+1} &= f(N_m,u_{j,m}) = f_{3,N} \left( \mathbf{f}_2 \left( \left[ \begin{array}{c} f_1(N_m)\\\sum_{j=1}^M u_{j,m} / M \end{array} \right] \right)  \right) \\
E_{m} &= h(N_m,u_{j,m}) = f_{3,E} \left( \mathbf{f}_2 \left( \left[ \begin{array}{c} f_1(N_m)\\\sum_{j=1}^M u_{j,m} / M \end{array} \right] \right)  \right) \ , \label{discrete_model_2}
\end{align}
\end{subequations}
where $f_1$ represents the pumping of inversion for a duration $t_{pump}$, $ \mathbf{f}_2$ the inversion and average intracavity power after the energy build-up for $t_{bu}$ and $f_{3,N}$ and $f_{3,E}$ the effect of coupling-out on inversion and pulse energy for $t_{coup}$. We will now show how such expressions can be derived from the continuous-time nonlinear model given by (17,20) of the paper, which we repeat here as
\begin{subequations} \label{CTNM}
\begin{align}
\begin{split}
\deriv{ N(t)}{t} & =   - b \gamma N(t) -  \sum_{j=1}^M\frac{b \lambda_j}{h c A_s}  \bigg( \exp \Big(\sigma(\lambda_j) N(t) -\alpha(\lambda_j)L \Big) -1 + \alpha(\lambda_j)L \bigg) P_j(t) \ + \\
& \frac{b \lambda_{p}}{h c A_p} \bigg( 1- \exp \Big(  \sigma_{p} \left((2 B_{45} + 1) N(t) -L N_{dop} \right) -\alpha_pL \Big) -\alpha_pL \bigg) P_p(t) \Bigg) \label{comp_red_rate} \end{split} \\
\deriv{P_j(t)}{t} &= \frac{2\sigma(\lambda_j)}{t_{RT}} N(t) P_j(t) - \frac{P_j(t)}{\tau(t,\lambda_j)} +  \frac{ 2}{t_{RT}} \Big( \frac{ 4 hc}{\lambda_j t_{RT}} \frac{\Delta\Omega}{4\pi} \mathfrak{C}(t,\lambda_j) +\alpha_{RS}L P_j(t) \Big) \ . \label{continuous_P_equation}
\end{align}
\end{subequations}
Since (\ref{CTNM}) cannot be solved analytically, we will proceed by solving simplified versions of (\ref{CTNM}) for the three sub-processes identified in Section 3.5 and then composing the respective solutions. At different stages of a switching cycle, different terms in (\ref{CTNM}) dominate, and hence, non-dominant ones will be neglected as a first simplification. Secondly, the spectrally distributed intracavity power $P_j(t)$ will be replaced by a spectral average $P(t)$. To further simplify the functional structure of (\ref{CTNM}), nonlinear depletion terms in (\ref{comp_red_rate}) will be used only to first or second order. For the effect of coupling-out on the population, a shape of $P(t)$ will be assumed in accordance with simulations to avoid solving (\ref{continuous_P_equation}) with time-varying $R(t)$. Finally, losses to $P(t)$ will be included by an operator-splitting approach.\\

\noindent \textbf{Pumping.} For the first sub-process, the dominant terms in (\ref{comp_red_rate}) are the ones including $P_p$ and $\gamma$ as the intracavity power $P_j$ is negligibly small for all $j$. Further assuming that
\begin{enumerate}
\item $\alpha_p=0$ (no pump losses within the active medium),
\item  $LN_{dop} \gg  N(t)$ (only a small fraction of all dopand atoms is in the excited state) and
\item  $P_p = $ const. (the pump power is not varied during a switching cycle),
\end{enumerate}
we have the differential equation
\begin{align*}
\deriv{ N(t)}{t} =  \frac{b \lambda_p}{hcA_p}P_p -b \gamma N(t)  \ ,
\end{align*}
which allows the simple solution
\begin{align}\label{new_simp_pump}
N(t_m+t_{pump}) =f_1( N_m)=  \frac{\lambda_p}{hcA_p \gamma} \left(1 - e^{-b\gamma t_{pump}} \right) P_p + N_m e^{-b\gamma t_{pump}} \ .
\end{align}
\textbf{Energy build-up.} The second sub-process is the energy build-up at high cavity quality $R(t) = R_{max}$. $N(t)$ and $P_j(t)$ start at $f_1( N_m)$ and $u_{j,m}:=P(t_m+t_{pump},\lambda_j) $, respectively. The dominant terms in (\ref{CTNM}) are now the depletion of $N(t)$ due to $P_j(t)$ in (\ref{comp_red_rate}) and the amplification of $P_j(t)$ as well as losses induced by $\tau(t,\lambda_j)$ in (\ref{continuous_P_equation}), which results in 
\begin{subequations}\label{DTNM_zwischen_1}
\begin{align}
\deriv{ N(t)}{t} & = - b \sum_{j=1}^M\frac{\lambda_j}{h c A_s}  \bigg( \exp \Big(\sigma(\lambda_j) N(t) -\alpha(\lambda_j)L \Big) -1 + \alpha(\lambda_j)L \bigg) P_j(t) \label{N_DTNM_zwischen_1} \\
\deriv{P_j(t)}{t} &= \frac{2\sigma(\lambda_j)}{t_{RT}} N(t) P_j(t) - \frac{P_j(t)}{\tau(t,\lambda_j)} \ . \label{P_DTNM_zwischen_1}
\end{align}
\end{subequations}
As mentioned above, the following steps will be taken to approximately solve (\ref{DTNM_zwischen_1}): 
\begin{enumerate}
\item $P_j(t)$ and $1/\tau(t,\lambda_j)$ are replaced by spectrally averaged quantities $P(t)$ and \linebreak
 $1/\bar{\tau} := \sum_{j=1}^M \left(2\alpha(\lambda_j)L-\ln(\eta(\lambda_j)R_{max})\right) / (M t_{RT})$ which describe the effective evolution of total energy within the cavity.
 \item The exponential in (\ref{N_DTNM_zwischen_1}) is Taylor expanded to second order such that after cancellations,
 \begin{subequations} \label{Taylor_approx_for_quadratic_eom}
\begin{align}
 &\exp \Big(\sigma(\lambda_j) N(t) -\alpha(\lambda_j)L \Big) -1 + \alpha(\lambda_j)L \approx \sigma(\lambda_j) N(t) + \frac{1}{2}  \Big(\sigma(\lambda_j) N(t) -\alpha(\lambda_j)L \Big)^2 \\
 & = \sigma(\lambda_j) N(t) (1-\alpha(\lambda_j)L) + \frac{1}{2 } (\sigma(\lambda_j) N(t))^2 + \frac{1}{2 } (\alpha(\lambda_j)L)^2 \ .
 \end{align}
 \end{subequations}
\item To find an approximate solution for $P(t)$, one needs to omit the quadratic terms in (\ref{Taylor_approx_for_quadratic_eom}) which represent second-order effects as usually, $\sigma(\lambda_j)N(t)< 1$ and $\alpha(\lambda_j)L \ll 1$. 
\item Losses are disregarded in (\ref{P_DTNM_zwischen_1}) and taken into account for $P(t)$ by including $1/\bar{\tau}$ through an operator-splitting approach.
\item The nonlinear depletion terms' effect on $N(t)$ is approximated by inserting the expression for $P(t)$ from the previous step into the simplified total rate equation from step 2.
\item The resulting expressions for $N(t)$ and $P(t)$ are finally augmented by fitting parameters $s_1$ and $s_2$ that mitigate modeling errors due to the simplifications performed above.
\end{enumerate}
After approximation steps 1 and 2 and with the auxiliary quantities 
\begin{align*}
q_0 &:= L^2 / 2 \sum_{j=1}^M \lambda_j \alpha^2(\lambda_j) \\
q_1 &:= \sum_{j=1}^M \lambda_j \sigma(\lambda_j) \left(1-\alpha(\lambda_j)L\right)\\
q_2 &:= 1 / 2 \sum_{j=1}^M \lambda_j \sigma^2(\lambda_j) \ ,
\end{align*}
the sums over $j$ in (\ref{N_DTNM_zwischen_1}) can be evaluated and one obtains
\begin{align}
\label{quadratic_eom_1}
\deriv{ N(t)}{t} = -\frac{b}{hcA_s} \left( q_0 + q_1 N(t) + q_2 N^2(t) \right) P(t) \ .
\end{align} 
If steps 3 and 4 (omission of nonlinear depletion and losses) are performed on (\ref{P_DTNM_zwischen_1}), both sides can be multiplied by $\lambda_j \left(1-\alpha(\lambda_j)L\right)$ and summed over $j$. This results in 
\begin{align*}
\Lambda \deriv{P(t)}{t} = \frac{2 q_1}{ t_{RT}}  N(t) P(t) \ ,
\end{align*}
with $\Lambda := \sum_{j=1}^M \lambda_j \left(1-\alpha(\lambda_j)L\right)$. Dividing by $\lambda$ yields 
\begin{align} \label{quadratic_eom_2}
\deriv{P(t)}{t} = \frac{2 q_1}{\Lambda t_{RT}}  N(t) P(t) \ .
\end{align}
For $N(t)$ and $P(t)$ starting the build-up at $N(t_m+t_{pump}$ and $u_m := \sum_{j=1}^M u_{j,m} / M$ with $q_2 N^2(t) \approx q_0 \approx 0$ by step 3, (\ref{quadratic_eom_1},\ref{quadratic_eom_2}) can be solved analytically, with the solution
\begin{subequations} \label{lin_eom_solution}
\begin{align}
&N_{lin}(t) = N(t_m+t_{pump}) \frac{a_N N(t_m+t_{pump}) + a_P u_m}{a_N N(t_m+t_{pump}) + a_P u_m \exp\left(\frac{2 b q_1 }{a_N a_P} \left(a_N N(t_m+t_{pump}) + a_P u_m\right)  t\right)} \\
&P_{lin}(t) = u_m \frac{a_N N(t_m+t_{pump}) + a_P u_m}{a_N N(t_m+t_{pump})\exp\left(-\frac{2 b q_1 }{a_N a_P} \left(a_N N(t_m+t_{pump}) + a_P u_m\right) t \right) + a_P u_m }  \ , \label{equ:P_lin}
\end{align}
\end{subequations}
and the auxiliary parameters
\begin{align*}
&a_N := 2h c A_s \\
&a_P := b \Lambda t_{RT}  \ .
\end{align*} 
Using an operator-splitting approach to finish the fourth step, we refine the solution $P_{lin}(t)$ and re-incorporate the losses $1/\bar{\tau}$ from (\ref{P_DTNM_zwischen_1}) via
\begin{align}\label{operator_splitting_equation}
\deriv{P(t)}{t} = -\frac{P(t)}{\bar{\tau}} \ .
\end{align} 
Composing the solution of this differential equation (exponential decay) with (\ref{equ:P_lin}), we get 
\begin{align}
P_{lin,OS}(t) = u_m \frac{a_N N(t_m+t_{pump}) + a_P u_m}{a_N N(t_m+t_{pump})\exp\left(-\frac{2 b q_1 }{a_N a_P} \left(a_N N(t_m+t_{pump}) + a_P u_m\right) t \right) + a_P u_m }  \exp\left( -t/\bar{\tau} \right) \ . \label{P_lin_OS}
\end{align}
As a next step (step 5), we can take the nonlinearity of (\ref{quadratic_eom_1}) due to $q_2$ and $q_0$ into account by plugging $P_{lin,OS}(t)$ from (\ref{P_lin_OS}) into (\ref{quadratic_eom_1}) instead of $P(t)$ and solving for $N(t)$. Without the $\bar{\tau}$ term, the approximate build-up solution for $N(t)$ reads as
\begin{align} \label{new_simp_buildup_1}
&N(t_m+t_{pump}+t_{bu})= \notag \\ 
&\frac{1}{2 q_2}\Big(  \textnormal{tanh} \Bigg(  \ln \left(  \frac{a_N N(t_m+t_{pump}) + a_P u_m) \exp\left(\frac{2 b q_1}{a_N a_P} \left(a_N N(t_m+t_{pump}) + a_P u_m\right) t_{bu}\right)}{a_N N(t_m+t_{pump}) + a_P u_m} \right) \times  \notag \\
& \times \frac{\sqrt{q_1^2-4 q_0 q_2}}{2 q_1} + \textnormal{atanh} \left( \frac{q_1+ 2q_2 N(t_m+t_{pump})}{\sqrt{q_1^2 - 4 q_0 q_2}} \right)   \Bigg) \sqrt{q_1^2-4 q_0 q_2} -q_1  \Big) \ .
\end{align}
Note that the solution exists for finite $\bar{\tau}$ as well. It is just a longer expression involving hypergeometric functions.\\
Simulations show that (\ref{new_simp_buildup_1}) fits depletion curves from more complex models relatively well. In a sixth step, the fit can be greatly improved, though, by incorporating a fitting parameter $s_2$ which accounts for nonlinearities beyond quadratic order in $N(t)$ in (\ref{N_DTNM_zwischen_1}). If $s_2$ is placed into (\ref{new_simp_buildup_1}) in 
\begin{align} \label{new_simp_buildup_1_with_s_2}
&N(t_m+t_{pump}+t_{bu})= \notag \\ 
&\frac{1}{2 q_2}\Big(  \textnormal{tanh} \Bigg(  \ln \left(  \frac{a_N N(t_m+t_{pump}) + a_P u_m) \exp\left(\frac{2 b q_1}{a_N a_P} \left(a_N N(t_m+t_{pump}) + a_P u_m\right)  s_2 t_{bu}\right)}{a_N N(t_m+t_{pump}) + a_P u_m} \right) \times  \notag \\
& \times \frac{\sqrt{q_1^2-4 q_0 q_2}}{2 q_1} + \textnormal{atanh} \left( \frac{q_1+ 2q_2 N(t_m+t_{pump})}{\sqrt{q_1^2 - 4 q_0 q_2}} \right)   \Bigg) \sqrt{q_1^2-4 q_0 q_2} -q_1  \Big) 
\end{align}
as an additional factor to $t_{bu}$, the quasi-exponential fall in population can be shifted in its time of occurrence to fit some desired behavior. \\
Shifting the time when $N(t)$ starts to be significantly depleted via the same $s_2$ improves the accuracy of (\ref{P_lin_OS}) as well. However, a second fitting parameter $s_1$ can be introduced to adjust how strongly the intracavity power rises, which is required due to disregarded nonlinear depletion terms from (\ref{N_DTNM_zwischen_1}) and the crude inclusion of losses via (\ref{operator_splitting_equation}). 
The approximate build-up solution for $P(t)$ is thus given by
\begin{align}
\label{new_simp_buildup_2_with_s_1} 
\begin{split} &P(t_m+t_{pump}+t_{bu}) =  \\
&u_m \left( \frac{a_N N(t_m+t_{pump}) + a_P u_m}{a_N N(t_m+t_{pump})\exp\left(-\frac{2 b q_1}{a_N a_P} \left(a_N N(t_m+t_{pump}) + a_P u_m)\right)  s_2 t_{bu} \right) + a_P u_m } \right)^{s_1} \ . \end{split}
\end{align}
In total, the energy build-up contributes to populations $N(t)$ and spectrally averaged intracavity powers $P(t)$ in (\ref{discrete_model_1},\ref{discrete_model_2}) according to
\begin{align*}
\mathbf{f}_2\left(\left[ \begin{array}{c}
N(t_m+t_{pump})\\
u_m
\end{array} \right]\right) = \left[\begin{array}{c}
N(t_m+t_{pump}+t_{bu})\\
P(t_m+t_{pump}+t_{bu})
\end{array}\right]
\end{align*}
with $N(t_m+t_{pump}+t_{bu})$ and $P(t_m+t_{pump}+t_{bu})$ from (\ref{new_simp_buildup_1_with_s_2}) and (\ref{new_simp_buildup_2_with_s_1}), respectively.\\

\noindent \textbf{Coupling-out.} Solving (\ref{CTNM}) with time-varying $R(t)$ and hence time-varying $\tau(\lambda_j,t)$ is not feasible. Therefore, for the third sub-process, 
\begin{enumerate}
\item steps 1 and 2 from the energy build-up are repeated, resulting in (\ref{quadratic_eom_1}).
\item It can be seen from simulations that for $R(t)$ falling from $R_{max}$ to $R_{min}$, $P(t)$ falls from some maximal value to approximately zero within a similar timeframe as well, while population is being further depleted. Thus, the shape of $P(t)$ during coupling-out is approximated by a line falling from $P(t_m+t_{pump}+t_{bu})$ to 0 within $t_{coup}$.
\end{enumerate}
These assumption entail that for the intracavity power, we can use the simple linear ansatz 
\begin{align} \label{equ:P_coupling_ansatz}
P(t) = P(t_m+t_{pump}+t_{bu}) \left( 1 - \left(t-(t_m+t_{pump}+t_{bu})\right)/t_{coup} \right) \ ,
\end{align}
which we again insert into (\ref{quadratic_eom_1}) and solve for $N(t)$.
Adding a third fitting parameter $s_3$ as a factor of $P(t)$, the population after coupling-out (here without the $\bar{\tau}$ term) is thus given by
\begin{align} \begin{split} \label{new_simp_coupling}
 N(t_{m+1}) & = N(t_m+t_{pump}+t_{bu}+t_{coup}) =  f_{3,N}\left(\left[ \begin{array}{c} N(t_m+t_{pump}+t_{bu})\\P(t_m+t_{pump}+t_{bu}) \end{array} \right] \right)\\
& = \frac{1}{2 q_2}\Big( \textnormal{tanh} \Bigg( \frac{b \sqrt{q_1^2-4q_0q_2}}{2 a_N} P(t_m+t_{pump}+t_{bu})s_3 t_{coup} \ + \\
& \textnormal{atanh} \left( \frac{2 q_2 N(t_m+t_{pump}+t_{bu}) + q_1}{\sqrt{q_1^2-4q_0q_2}} \right) \Bigg) \sqrt{q_1^2-4q_0q_2} - q_1 \Big) \ . \end{split}
\end{align}
The purpose of $s_3$ is to account for the fact that $P(t)$ is often not yet at its maximum at the moment the cavity is opened and instead continues to further rise for a short time.\\

\noindent The pulse-to-pulse population dynamics as given by (\ref{discrete_model_1}) can finally be obtained by composing the sub-process solutions (\ref{new_simp_pump}), (\ref{new_simp_buildup_1_with_s_2},\ref{new_simp_buildup_2_with_s_1}) and (\ref{new_simp_coupling}). \\

\noindent \textbf{Output pulse energy.} We are also interested in finding a simple representation of the output energy as given by (\ref{discrete_model_2}). The most relevant contribution to this comes from output coupling. Hence, we again use the spectrally averaged intracavity power ansatz from (\ref{equ:P_coupling_ansatz}) with (\ref{new_simp_buildup_2_with_s_1}) as well as $N(t_m+t_{pump}+t_{bu})$ from (\ref{new_simp_buildup_1}) and insert them into the deterministic version of the output power given by (21) from the paper. Finally, we convert this to the approximate $m$-th output energy by taking the integral over time while $R(t)$ is falling, i.e. $t_m+t_{pump}+t_{bu} \leq t < t_m+t_{pump}+t_{bu}+t_{close}$, and when $R(t)$ has fallen to its minimum value $R_{min}$, i.e. between $t_{close}$ and $t_{coup}$.  The formula for this reads as
\begin{align*} 
E_m &= f_{3,E}\left(  \left[ \begin{array}{c} N(t_m + t_{pump}+t_{bu})\\ P(t_m+t_{pump}+t_{bu}) \end{array} \right] \right) = \sum_{j=1}^M \int_{t_m+t_{pump}+t_{bu}}^{t_m+t_{pump}+t_{bu}+t_{coup}} P_{out}(t,\lambda_j) \ \textnormal{d}t \approx  \\
&\approx \sum_{j=1}^M \int_{t_m+t_{pump}+t_{bu}}^{t_m+t_{pump}+t_{bu}+t_{coup}} \frac{\left( 1-R(t) \right)}{R(t)} \times\\ 
& \times  \left(  \frac{P(t)\exp\left( \sigma(\lambda_j)N(t_m+t_{pump}+t_{bu})-\alpha(\lambda_j)L \right) +  2 \frac{hc}{\lambda_j t_{RT}} \frac{\Delta\Omega}{4\pi} \sigma(\lambda_j)N(t_m+t_{pump}+t_{bu})}{\frac{1}{\sqrt{\eta(\lambda_j)}R(t)} + \exp\left( \sigma(\lambda_j)N(t_m+t_{pump}+t_{bu})-\alpha(\lambda_j)L \right) }  \right) \ \textnormal{d}t \ ,
\end{align*}
with 
\begin{align*}
R(t) &= R_{max} - (R_{max} - R_{min})\frac{t-\left(t_m+t_{pump}+t_{bu}\right)}{t_{close}} \ \textnormal{for} \ t_m+t_{pump}+t_{bu} \leq t < t_m+t_{pump}+t_{bu}+t_{close} \\
R(t) &= R_{min}  \ \textnormal{for} \ t_m+t_{pump}+t_{bu} +t_{close} \leq t < t_m+t_{pump}+t_{bu}+t_{coup} = t_{m+1} \\
P(t) &= P(t_m+t_{pump}+t_{bu}) \left( 1 - \left(t-(t_m+t_{pump}+t_{bu})\right)/t_{coup} \right)  \ \textnormal{for} \ t_m+t_{pump}+t_{bu} \leq t <  t_{m+1} \ .
\end{align*}
The integral can be carried out analytically, which symbolically endows us with $f_{3,E}$ and thus completes the derivation of (\ref{discrete_model_1},\ref{discrete_model_2}).

\section{Simulation parameters for Nd:YAG laser}

\begin{table}[H]
\begin{center}
\begin{tabular}{c | c}
$\lambda_p$ & 806 nm \\
$\lambda_A$ & 946 nm \\
$\lambda_B$ & 1064 nm \\
$A_p$ 		& 3.526 $\times 10^{-7}$ m$^2$ \\
$A_s$ 		& 4.646 $\times 10^{-7}$ m$^2$ \\
bandwidth $\sigma_A$ & 0.99 nm \\
bandwidth $\sigma_B$ & 0.60 nm \\
maximum $\sigma_A$ & 5 $\times 10^{-24}$ m$^2$ \\
maximum $\sigma_B$ & 28 $\times 10^{-24}$ m$^2$ \\
$L$ & 12 mm \\
$N_{dop}$ & 6.9 $\times 10^{25}$ m$^{-3}$ \\
$\gamma_{42}$ & 4545 s$^{-1}$ \\
$\gamma_{43}$ & 4348 s$^{-1}$ \\
$\gamma_{54}$ & 4$\times 10^8$ s$^{-1}$ \\
$\gamma_{31}$ & 4$\times 10^7$ s$^{-1}$ \\
$\gamma_{21}$ & 4$\times 10^8$ s$^{-1}$ \\
$\sigma_p$ & 7.7 $\times 10^{-24}$ m$^2$ \\
$B_{45}$ &  0.0166\\
$B_{12}$ &  0.0133\\
$B_{13}$ &  3.982 $\times 10^{-5}$\\
$\gamma_{45}$ & $B_{45}$ $\gamma_{54}$ \\
$\gamma_{13}$ & $B_{13}$ $\gamma_{31}$ \\
$\gamma_{12}$ & $B_{12}$ $\gamma_{21}$ \\
$t_{RT}$ &  1.4357 nm \\
$\eta(\lambda_j)$ & around 0.8, centered at $\lambda_B$ \\
$\alpha_p$ & 0 m$^{-1}$ \\
$\alpha(\lambda_j)$ & 0 m$^{-1}$ \\
$\alpha_{RS}$ & 0 m$^{-1}$ \\
$\Delta\Omega$ & 2$\pi$\\
$b$ & $\frac{1}{1+B_{45}}$\\
$\gamma$ & $\gamma_{42}$ + $\gamma_{43}$

\end{tabular}
\end{center}
\end{table}

%
%
%
%
%

\bibliography{local}{}
\bibliographystyle{plain}

\end{document}